\documentclass[12pt]{article}
\usepackage{amsmath}
\usepackage{amssymb}
\usepackage{amsfonts}
\usepackage{amscd}
\usepackage[all]{xy}

\renewcommand{\theequation}{\arabic{section}.\arabic{equation}}

\newcommand{\level}{\thicklines\line(1,0){80}}
\newcommand{\ord}{\thicklines\line(1,0){2}}
\newcommand{\trait}{\line(0,1){2}}
\newcommand{\fleche}{\thicklines\vector(0,1){24}}
\newcommand{\niveau}{\thicklines\line(1,0){15}}
\newcommand{\arrow}{\thicklines\vector(1,0){8}}

\newsavebox{\dash}

\savebox{\dash}(5,20)[b]{\multiput(0,0)(0,4){3}{\trait}}

\title{
Combined state-adding and state-deleting approaches to type III multi-step rationally-extended potentials: 
applications to ladder operators and superintegrability}

\author{Ian Marquette$^{1,}$\thanks{Electronic address: i.marquette@uq.edu.au} \ and Christiane Quesne$^{2,}$\thanks{Electronic address: cquesne@ulb.ac.be}\\
{\small\sl $^1$ School of Mathematics and Physics, The University of Queensland,}\\
{\small \sl Brisbane, QLD 4072, Australia}\\
{\small\sl $^2$ Physique Nucl\'eaire Th\'eorique et Physique Math\'ematique, 
Universit\'e Libre de Bruxelles,} \\ 
{\small \sl Campus de la Plaine CP229, Boulevard~du Triomphe, B-1050
Brussels, Belgium}}
\date{ }
\begin{document}
\baselineskip=22pt plus 1pt minus 1pt
\maketitle

\begin{abstract}
Type III multi-step rationally-extended harmonic oscillator and radial harmonic oscillator potentials, characterized by a set of $k$ integers $m_1$, $m_2$, \ldots, $m_k$, such that $m_1 < m_2 < \cdots < m_k$ with $m_i$ even (resp.\ odd) for $i$ odd (resp.\ even), are considered. The state-adding and state-deleting approaches to these potentials in a supersymmetric quantum mechanical framework are combined to construct new ladder operators. The eigenstates of the Hamiltonians are shown to separate into $m_k+1$ infinite-dimensional unitary irreducible representations of the corresponding polynomial Heisenberg algebras. These ladder operators are then used to build a higher-order integral of motion for seven new infinite families of superintegrable two-dimensional systems separable in cartesian coordinates. The finite-dimensional unitary irreducible representations of the polynomial algebras of such systems are directly determined from the ladder operator action on the constituent one-dimensional Hamiltonian eigenstates and provide an algebraic derivation of the superintegrable systems whole spectrum including the level total degeneracies.
\end{abstract}

\vspace{0.5cm}


 
\newpage
%
%
\section{INTRODUCTION}

In one-dimensional quantum mechanics, there has been a continuing interest in constructing exactly solvable potentials. 
For such a purpose, supersymmetric quantum mechanics (SUSYQM) \cite{cooper} (combined with the concept of (translationally)
 shape invariant potentials (SIP) \cite{genden}) has provided a very powerful tool consistent with the Schr\"odinger factorization
 method \cite{schrodinger, infeld} and the Darboux intertwining one \cite{darboux}. On using first-order intertwining operators 
expressed in terms of some nodeless seed solution of a SIP Schr\"odinger equation, one can indeed build a partner potential, whose 
spectrum may differ by at most one level from that of the initial one. The ground-state energy is removed if the seed solution is the
 ground-state wavefunction of the latter (state-deleting case). If the seed solution has an energy below the ground state, it gives rise
 to an extra bound state below the initial spectrum (state-adding case) or it leads to the same bound-state spectrum (isospectral case) 
according to whether its inverse is normalizable or not.\par
%
%
More flexibility may be achieved by resorting to $n$th-order intertwining operators with $n>1$, constructed
 from $n$ seed solutions of the initial Hamiltonian Schr\"odinger equation \cite{andrianov93, andrianov95a, andrianov95b, bagrov, samsonov96, samsonov99, bagchi99, aoyama, fernandez04}. 
On the mathematical side, this corresponds to the Crum extension \cite{crum} of the Darboux method. Crum treated more specifically the example of 
seed solutions that are consecutive wavefunctions starting from the ground-state one, which is the simplest case considered in higher-order SUSYQM.
 Later on, Krein \cite{krein} and Adler \cite{adler} independently generalized the Crum results to chains of bound-state wavefunctions that may be 
lacunary with even gaps. This modification enriched the possibilities offered by higher-order SUSYQM at its very beginning \cite{bagrov, samsonov96},
 but along the years the latter has continued developing itself as a very efficient tool for spectral design, i.e., for building quantum systems with
 desired features (for a recent review see Ref.~\cite{andrianov12}).\par
%
%
On the other hand, during the last few years, a lot of research activity has been devoted to the construction of exceptional orthogonal
 polynomials (EOP) \cite{gomez10a, gomez09, gomez10b, gomez12a, gomez12b, gomez13a, gomez13b, gomez13c, fellows, cq08, bagchi09, cq09, cq11a, cq11b, cq12a, cq12b, marquette13a, odake09, odake10, sasaki, odake11, odake13a, odake13b, grandati11a, grandati11b, grandati12a, grandati12b, grandati13, ho11a, ho11b}, 
which are new complete and orthogonal polynomial systems $X_m$ extending the classical families of Hermite, Laguerre, and Jacobi \cite{erdelyi}.
 In contrast with the latter, the former admit some gaps in the sequence of their degrees, the total number of them being referred to as the 
codimension $m$.\par
%
%
The first examples of EOP, the so-called Laguerre and Jacobi $X_1$ families, were proposed in the context of Sturm-Liouville theory \cite{gomez10a, gomez09}, 
but it soon appeared that, apart from some gauge factor, they constituted the main part of the bound-state wavefunctions of some rationally-extended SIP, 
which were themselves shape invariant \cite{cq08}. From that time onwards, it has proved useful to adopt a SUSYQM or Darboux approach to the 
problem \cite{bagchi09, cq09}. EOP of arbitrarily large codimension $m$ have then been constructed \cite{odake09, odake10}. A further extension 
has been the appearance of multi-indexed families $X_{m_1, m_2, \ldots, m_k}$ \cite{gomez12a, odake11} in the context of higher-order SUSYQM or its variants.
 All of these EOP turn out to be associated with some solvable rational extensions of well-known SIP.\par
%
%
{}From the very beginning \cite{cq09}, three different classes of $X_m$ EOP have been identified. Later on, they were shown to be connected with discrete symmetries of the starting SIP \cite{grandati11a}. The rationally-extended SIP corresponding to type I or type II EOP turn out to be strictly isospectral to the starting potential and to be shape invariant as the latter. In contrast, those associated with type III EOP (which are the only ones in the case of Hermite EOP) have an extra bound state below the starting potential spectrum and are not shape invariant. In the multi-indexed case, combining these three types may therefore in general lead to a rich variety of solvable rationally-extended potentials. In some recent works \cite{gomez13c, odake13a}, special attention has been paid to higher-order (pure) type III extensions. It has been shown, in particular, that, up to some energy shift, adding $k$ new states below a SIP energy spectrum is equivalent to deleting some (appropriately chosen) excited states, thereby generalizing previous observations \cite{oblomkov, felder}.\par
%
%
Going from one to two dimensions, some superintegrable systems related to EOP have been recently considered \cite{post12, marquette13b, marquette13c}. Let us recall that a two-dimensional superintegrable quantum Hamiltonian system $H$ is characterized by the existence of two integrals of motion $X$ and $Y$, which are well-defined operators and form with $H$ an algebraically independent set (for a recent review see \cite{miller}). The best known of them are the quadratically superintegrable systems, i.e., those allowing two (at most) second-order integrals of motion. Their study began in the mid 1960s \cite{winternitz} and by now they have been completely classified in conformally flat spaces \cite{kalnins01, kalnins03, kalnins05a, kalnins05b, kalnins06, daska01, daska06, ballesteros, cq11c, post11}. In the higher-order case, the direct approach for determining integrals of motion becomes more and more difficult as their order increases, as it has recently been shown for third \cite{gravel02, gravel04, marquette09a, marquette09b} and quartic order \cite{marquette13d}. For such a reason, some other approaches, based on ladder operators \cite{marquette10}, recurrence relations \cite{kalnins11, kalnins12}, or SUSYQM \cite{marquette09c, demir}, have been proposed. As superintegrable systems related to EOP belong to such a higher-order case, the recurrence relation and the ladder operator methods have been applied in Ref.~\cite{post12} and in Refs.~\cite{marquette13b, marquette13c}, respectively.\par
%
%
In Ref.~\cite{marquette13b}, some two-dimensional systems connected with type III Hermite EOP, as well as to
type I, II, or III Laguerre EOP, were analyzed. The results proved entirely satisfactory for type I or II, but for type III it was not possible to derive the whole energy spectrum from the representations of the polynomial algebra generated by the integrals of motion. For the constituent one-dimensional Hamiltonians related to EOP, this study used ladder operators obtained by combining those for the partner SIP with the supercharges of the SUSYQM approach. Such a type of ladder operators is a well-known one and the polynomial Heisenberg algebras (PHA) they close with the Hamiltonian have been the topic of several studies \cite{fernandez04, fernandez99, carballo}.\par
%
%
With the aim of providing an adequate approach to superintegrable systems connected with type III EOP, some novel ladder operators were introduced in Ref.~\cite{marquette13c} for the case of a rationally-extended harmonic oscillator derived in first-order SUSYQM and related to type III Hermite EOP.\par
%
%
The purpose of the present paper is twofold: first to propose a generalization of the latter ladder operators for type III multi-step rationally-extended harmonic oscillator or radial harmonic oscillator potentials, based on the (essential) equivalence of the state-adding and state-deleting approaches referred to above \cite{gomez13c, odake13a}, and second to use them for constructing new superintegrable systems, extending some of those considered in Ref.~\cite{marquette13b}, as well as for providing an algebraic derivation of their spectrum.\par
%
%
In Sec.~II, the state-adding and state-deleting approaches to type III multi-step rationally-extended harmonic oscillator and radial harmonic oscillator are briefly reviewed. In Sec.~III, they are combined to construct new ladder operators for such extensions. The PHA generated by these ladder operators and their unitary irreducible representations (unirreps) are also determined. In Sec.~IV, from the one-dimensional systems considered in Secs.~II and III, some new two-dimensional superintegrable systems are built and the polynomial algebras generated by their integrals of motion are obtained. The finite-dimensional unirreps of the latter are also determined from the known action of the ladder operators on the eigenstates of the constituent Hamiltonians, thereby providing an algebraic derivation of the superintegrable system spectrum. Finally, Sec.~V contains the conclusion.\par
%
%
\section{STATE ADDING VERSUS STATE DELETING FOR TYPE III MULTI-STEP RATIONALLY-EXTENDED POTENTIALS}

In $n$th-order SUSYQM \cite{andrianov93, andrianov95a, andrianov95b, bagrov, samsonov96, samsonov99, bagchi99, aoyama, fernandez04}, the two partner Hamiltonians
\begin{equation}
  H^{(i)} = - \frac{d^2}{dx^2} + V^{(i)}(x), \qquad i=1, 2,
\end{equation}
intertwine with two $n$th-order differential operators $\cal A$ and ${\cal A}^{\dagger}$ as
\begin{equation}
  {\cal A} H^{(1)} = H^{(2)} {\cal A}, \qquad {\cal A}^{\dagger} H^{(2)} = H^{(1)} {\cal A}^{\dagger}.
  \label{eq:intertwine}
\end{equation}
We consider here the case where the latter can be factorized as
\begin{equation}
  {\cal A} = A^{(n)} \cdots A^{(2)} A^{(1)}, \qquad {\cal A}^{\dagger} = A^{(1)\dagger} A^{(2)\dagger} \cdots
  A^{(n)\dagger},
\end{equation}
into products of $n$ first-order differential operators
\begin{equation}
  A^{(i)} = \frac{d}{dx} + W^{(i)}(x), \qquad A^{(i)\dagger} = - \frac{d}{dx} + W^{(i)}(x), \qquad i=1, 2, \ldots, n, 
\end{equation}
with real superpotentials $W^{(i)}(x)$, $i=1$, 2, \ldots, $n$. Such superpotentials, which are assumed to be obtained from $n$ different seed solutions $\varphi_i(x)$, $i=1$, 2, \ldots, $n$, of the Schr\"odinger equation associated with $H^{(1)}$, can be written as \cite{fernandez04, crum}
\begin{equation}
  W^{(i)}(x) = - \frac{d}{dx} \log \varphi^{(i)}(x), \qquad i=1, 2, \ldots, n,
\end{equation}
with
\begin{equation}
  \varphi^{(1)}(x) = \varphi_1(x), \qquad \varphi^{(i)}(x) = \frac{{\cal W}(\varphi_1, \varphi_2, \ldots, \varphi_i)}
  {{\cal W}(\varphi_1, \varphi_2, \ldots, \varphi_{i-1})}, \qquad i=2, 3, \ldots, n.
\end{equation}
Here ${\cal W}(\varphi_1, \varphi_2, \ldots, \varphi_i)$ denotes the Wronskian of $\varphi_1(x)$, $\varphi_2(x)$, \ldots, $\varphi_i(x)$ \cite{muir}. Since the potentials of the two partner Hamiltonians are linked by the relationship \cite{fernandez04, crum}
\begin{equation}
  V^{(2)}(x) = V^{(1)}(x) - 2 \frac{d^2}{dx^2} \log {\cal W}(\varphi_1, \varphi_2, \ldots, \varphi_n),
  \label{eq:partner}
\end{equation}
a regular potential $V^{(2)}(x)$ is obtained provided the Wronskian ${\cal W}(\varphi_1, \varphi_2, \ldots, \varphi_n)$ does not vanish on the defining interval of $V^{(1)}(x)$.\par
%
%
\subsection{Harmonic oscillator case}

If $V^{(1)}(x)$ is the harmonic oscillator potential
\begin{equation}
  V^{(1)}(x) = x^2, \qquad - \infty < x < \infty,  \label{eq:HO}
\end{equation}
then the starting Hamiltonian $H^{(1)}$ has an infinite number of bound-state wavefunctions
\begin{equation}
  \psi^{(1)}_{\nu}(x) \propto \psi_{\nu}(x) = H_{\nu}(x) e^{-\frac{1}{2} x^2}, \qquad \nu=0, 1, 2, \ldots,
  \label{eq:wf-1}
\end{equation}
with corresponding energies
\begin{equation}
  E^{(1)}_{\nu} = 2\nu+1, \qquad \nu=0, 1, 2, \ldots.  \label{eq:energy-HO}
\end{equation}
Here $H_{\nu}(x)$ denotes a $\nu$th-degree Hermite polynomial \cite{erdelyi}.\par
%
%
This Hamiltonian also possesses raising and lowering operators $a^{\dagger}$ and $a$, which are first-order differential operators given by
\begin{equation}
  a=\frac{d}{dx}+x,\qquad a^{\dagger}=-\frac{d}{dx}+x. \label{eq:ladder-HO}
\end{equation}
They satisfy the Heisenberg algebra of the form
\begin{equation}
\begin{split}
  & [H^{(1)}, a^{\dagger}] = 2 a^{\dagger}, \qquad [H^{(1)}, a] = - 2 a, \\
  & [a, a^{\dagger}] = Q(H^{(1)}+2) - Q(H^{(1)}), 
\end{split}  \label{eq:PHA-HO}
\end{equation}
with
\begin{equation}
   Q(H^{(1)})=H^{(1)}-1. \label{eq:Q-HO}
\end{equation}
The action of the lowering operator on the wavefunctions is given by
\begin{equation}
\begin{split}
  & a\psi^{(1)}_{\nu} = 0, \qquad \nu=0,  \\
  & a\psi^{(1)}_{\nu} = [\nu]^{1/2} \psi^{(1)}_{\nu-1}, \qquad \nu=1, 2, \ldots. 
\end{split}  \label{eq:action-HO}
\end{equation}
\par
%
%
In the state-adding case, let us choose $n=k$ seed functions among the polynomial-type eigenfunctions $\phi_m(x)$ of $H^{(1)}$ below the ground-state energy $E^{(1)}_0$. These eigenfunctions, associated with the eigenvalues $E_m = - 2m -1$, can be written as \cite{bagrov, fellows, marquette13a}
\begin{equation}
  \phi_m(x) = {\cal H}_m(x) e^{\frac{1}{2} x^2}, \qquad m=0, 1, 2, \ldots,  \label{eq:phi}
\end{equation}
where ${\cal H}_m(x) = (-{\rm i})^m H_m({\rm i}x)$ is a $m$th-degree pseudo-Hermite polynomial (called twisted Hermite polynomial in the mathematical literature \cite{kwon}). For even $m$ values, the eigenfunctions $\phi_m(x)$ are nodeless on the whole real line, while for odd $m$ ones, they have a single zero at $x=0$.\par
%
%
With the choice $(\varphi_1, \varphi_2, \ldots, \varphi_n) \to (\phi_{m_1}, \phi_{m_2}, \ldots, \phi_{m_k})$, the partner potential (\ref{eq:partner}) of (\ref{eq:HO}) turns out to be nonsingular if $m_1 < m_2 < \cdots < m_k$ with $m_i$ even (resp.\ odd) for $i$ odd (resp.\ even) \cite{bagrov, samsonov96, fernandez04}. On using Eq.~(\ref{eq:phi}) and standard properties of Wronskians \cite{muir}, we can write ${\cal W}(\phi_{m_1}, \phi_{m_2}, \ldots, \phi_{m_k}) = \exp(\frac{1}{2}kx^2) {\cal W}({\cal H}_{m_1}, {\cal H}_{m_2}, \ldots, {\cal H}_{m_k})$, so that $V^{(2)}(x)$ becomes
\begin{equation}
  V^{(2)}(x) = x^2 - 2k - 2 \frac{d^2}{dx^2} \log {\cal W}({\cal H}_{m_1}, {\cal H}_{m_2}, \ldots, 
  {\cal H}_{m_k}).  \label{eq:partner-add}
\end{equation}
Its spectrum is given by
\begin{equation}
  E^{(2)}_{\nu} = 2\nu + 1, \qquad \nu = -m_k-1, \ldots, -m_2-1, -m_1-1, 0, 1, 2, \ldots, \label{eq:energy-add}
\end{equation}
and the corresponding wavefunctions are \cite{fernandez04}
\begin{equation}
\begin{split}
  & \psi^{(2)}_{\nu}(x) \propto \frac{{\cal W}(\phi_{m_1}, \phi_{m_2}, \ldots, \phi_{m_k}, \psi_{\nu})}
       {{\cal W}(\phi_{m_1}, \phi_{m_2}, \ldots, \phi_{m_k})}, \qquad \nu=0, 1, 2, \ldots, \\
  & \psi^{(2)}_{-m_i-1}(x) \propto \frac{{\cal W}(\phi_{m_1}, \phi_{m_2}, \ldots, \check{\phi}_{m_i}, \ldots,
       \phi_{m_k})}{{\cal W}(\phi_{m_1}, \phi_{m_2}, \ldots, \phi_{m_k})}, \qquad i=1, 2, \ldots, k.
\end{split}
\end{equation}
Here ${\cal W}(\phi_{m_1}, \phi_{m_2}, \ldots, \check{\phi}_{m_i}, \ldots, \phi_{m_k})$ means that $\phi_{m_i}(x)$ is excluded from the Wronskian for $k>1$, while the latter reduces to one for $i=k=1$.\par
%
%
To get (essentially) equivalent results for the two partner potentials in the state-deleting case \cite{gomez13c, odake13a}, we have to take (at least) $n = m_k + 1 - k$ bound-state wavefunctions of $H^{(1)}$ as seed functions: $(\varphi_1, \varphi_2, \ldots, \varphi_n) \to (\psi_1, \psi_2, \ldots, \check{\psi}_{m_k-m_{k-1}}, \ldots, \check{\psi}_{m_k-m_2}, \ldots, \check{\psi}_{m_k-m_1}, \ldots, \psi_{m_k})$ \cite{footnote}. The latter will then be suppressed from the spectrum. To distinguish this case from the previous one, let us put a bar above all corresponding quantities. Hence, the two partner Hamiltonians are now denoted by $\bar{H}^{(i)}$, $i=1$, 2, and the associated potentials are
\begin{equation}
  \bar{V}^{(1)}(x) = V^{(1)}(x) = x^2  \label{eq:V-bar-1}
\end{equation}
and $\bar{V}^{(2)}(x)$ given by Eq.~(\ref{eq:partner}) with the appropriate choice for $(\varphi_1, \varphi_2, \ldots, \varphi_n)$. Equation (\ref{eq:wf-1}), combined with standard properties of Wronskians, leads to ${\cal W}(\psi_1, \psi_2, \ldots, \check{\psi}_{m_k-m_{k-1}}, \ldots, \check{\psi}_{m_k-m_1}, \ldots, \psi_{m_k}) = \exp[- \frac{1}{2} (m_k+1-k) x^2]$ ${\cal W}(H_1, H_2, \ldots, \check{H}_{m_k-m_{k-1}}, \ldots, \check{H}_{m_k-m_1}, \ldots, H_{m_k})$, thence
\begin{equation}
  \bar{V}^{(2)}(x) = x^2 + 2(m_k+1-k) - 2 \frac{d^2}{dx^2} \log {\cal W}(H_1, H_2, \ldots, 
  \check{H}_{m_k-m_{k-1}}, \ldots, \check{H}_{m_k-m_1}, \ldots, H_{m_k}). \label{eq:partner-del}
\end{equation}
This potential is nonsingular  provided the gaps between the surviving levels with $\nu = 0, m_k-m_{k-1}, \ldots, m_k-m_2, m_k-m_1, m_k+1, m_k+2, \ldots$, correspond to even numbers of consecutive levels \cite{krein, adler}. This leads to the same conditions on $m_1$, $m_2$, \ldots, $m_k$ as in the state-adding case. The resulting spectra are now
\begin{equation}
\begin{split}
  & \bar{E}^{(1)}_{\nu} = 2\nu+1, \qquad \nu=0, 1, 2, \ldots, \\
  & \bar{E}^{(2)}_{\nu} = 2m_k + 2\nu + 3, \qquad \nu=-m_k-1, \ldots, -m_2-1, -m_1-1, 0, 1, 2, \ldots,
\end{split}  \label{eq:energy-del}
\end{equation} 
with corresponding wavefunctions $\bar{\psi}^{(1)}_{\nu}(x) = \psi^{(1)}_{\nu}(x)$, $\nu=0$, 1, 2, \ldots, and
\begin{equation}
\begin{split}
  & \bar{\psi}^{(2)}_{\nu}(x) \propto \frac{{\cal W}(\psi_1, \psi_2, \ldots, \check{\psi}_{m_k-m_{k-1}}, \ldots,
       \check{\psi}_{m_k-m_1}, \ldots, \psi_{m_k}, \psi_{m_k+1+\nu})}{{\cal W}(\psi_1, \psi_2, \ldots, 
       \check{\psi}_{m_k-m_{k-1}}, \ldots, \check{\psi}_{m_k-m_1}, \ldots, \psi_{m_k})}, \\
  & \qquad \nu=0, 1, 2, \ldots, \\
  & \bar{\psi}^{(2)}_{-m_i-1}(x) \propto \frac{{\cal W}(\psi_1, \psi_2, \ldots, \check{\psi}_{m_k-m_{k-1}}, \ldots,
       \psi_{m_k-m_i}, \ldots, \check{\psi}_{m_k-m_1}, \ldots, \psi_{m_k})}{{\cal W}(\psi_1, \psi_2, \ldots, 
       \check{\psi}_{m_k-m_{k-1}}, \ldots, \check{\psi}_{m_k-m_1}, \ldots, \psi_{m_k})}, \\
  & \qquad i=1, 2, \ldots, k.    
\end{split}  \label{eq:partner-wf}
\end{equation}
\par
%
%
Since it has been proved that the two Wronskians in Eqs.~(\ref{eq:partner-add}) and (\ref{eq:partner-del}) only differ by some multiplicative constant \cite{odake13a}, the two obtained potentials $V^{(2)}(x)$ and $\bar{V}^{(2)}(x)$ are the same up to some additive constant:
\begin{equation}
  V^{(2)}(x) + 2m_k + 2 = \bar{V}^{(2)}(x).  \label{eq:V-bar-2}
\end{equation}
This result agrees with the respective energies, given in Eqs.~(\ref{eq:energy-add}) and (\ref{eq:energy-del}), and also implies that $\bar{\psi}^{(2)}_{\nu}(x) \propto \psi^{(2)}_{\nu}(x)$, $\nu = -m_k-1$, \ldots, $-m_2-1$, $-m_1-1$, 0, 1, 2, \ldots.\par
%
%
\subsection{Radial harmonic oscillator case}

The radial harmonic oscillator case is more complicated than that of the linear one because the potential $V_l(x) = \frac{1}{4} x^2 + \frac{l(l+1)}{x^2}$, $0 < x < \infty$, now depends on a parameter $l$ (the angular momentum quantum number), which is changed in SUSYQM transformations.\par
%
%
So, in the state-adding case, to get a partner potential that is a rational extension of $V_l(x)$, we have to start from
\begin{equation}
\begin{split}
  V^{(1)}(x) &= V_{l+k}(x) = \frac{1}{4} x^2 + \frac{(l+k)(l+k+1)}{x^2}, \qquad 0 < x < \infty, \\
  &= \frac{1}{2} z + \frac{(2\alpha + 2k -1)(2\alpha + 2k +1)}{8z}, \qquad 0 < z < \infty, 
\end{split}  \label{eq:RHO}
\end{equation}
with $z \equiv \frac{1}{2} x^2$, $\alpha \equiv l + \frac{1}{2}$. The bound-state energies and wavefunctions of the starting Hamiltonian $H^{(1)}$ are therefore
\begin{equation}
  E^{(1)}_{l+k,\nu} = 2\nu + \alpha + k + 1, \qquad \nu=0, 1, 2, \ldots,  \label{eq:RHO-E}
\end{equation}
and
\begin{equation}
\begin{split}
  \psi^{(1)}_{l+k,\nu}(x) &\propto \psi^{(l+k)}_{\nu}(x) = \eta_{l+k}(z) L^{(\alpha+k)}_{\nu}(z) \\
  &\propto x^{l+k+1} e^{-\frac{1}{4}x^2} L^{(l+k+\frac{1}{2})}_{\nu}(\tfrac{1}{2}x^2), \qquad \nu=0, 1, 2, \ldots,
\end{split}
\end{equation}
with
\begin{equation}
  \eta_l(z) = z^{\frac{1}{4}(2\alpha+1)} e^{-\frac{1}{2}z}.
\end{equation}
Here $L^{(\alpha)}_{\nu}(z)$ denotes a $\nu$th-degree Laguerre polynomial \cite{erdelyi}. The $n=k$ seed functions are chosen among the polynomial-type eigenfunctions of $H^{(1)}$,
\begin{equation}
\begin{split}
  \phi^{(l+k)}_m(x) &= \bigl(\eta_{l+k-1}(z)\bigr)^{-1} L^{(-\alpha-k)}_m(-z) \propto x^{-l-k} e^{\frac{1}{4}x^2}
      L^{(-l-k-\frac{1}{2})}_m(-\tfrac{1}{2}x^2), \\
  & \quad m=0, 1, 2, \ldots,
\end{split}      
\end{equation}
at an energy $E^{(l+k)}_m = \alpha + k - 2m - 1$ below the ground-state energy $E^{(1)}_{l+k,0}$ \cite{cq09, odake13a, grandati11a}.\par
%
%
This Hamiltonian also possesses raising and lowering operators $a^{\dagger}$ and $a$, which are second-order differential operators given by
\begin{equation}
\begin{split}
  & a= \frac{1}{4}\left( 2 \frac{d^2}{dx^2}+2x\frac{d}{dx}+\frac{1}{2}x^2-\frac{2(l+k)(l+k+1)}{x^2}+1\right),   \\
  & a^{\dagger}=\frac{1}{4}\left( 2 \frac{d^2}{dx^2}-2x\frac{d}{dx}+\frac{1}{2}x^2-\frac{2(l+k)(l+k+1)}{x^2}-1\right),
\end{split}  \label{eq:ladder-RHO}
\end{equation}
and satisfy a polynomial Heisenberg algebra of the form
\begin{equation}
\begin{split}
  & [H^{(1)}, a^{\dagger}] = 2 a^{\dagger}, \qquad [H^{(1)}, a] = - 2 a, \\
  & [a, a^{\dagger}] = Q(H^{(1)}+2) - Q(H^{(1)}), 
\end{split}  \label{eq:PHA-RHO}
\end{equation}
with
\begin{equation}
 Q(H^{(1)})= \frac{1}{16}\left( 2 H^{(1)}-3-2l-2k\right)\left( 2 H^{(1)}-1+2l+2k\right). \label{eq:Q-RHO}
\end{equation}
The action of the lowering operator on the wavefunctions is given by
\begin{equation}
\begin{split}
  & a\psi^{(1)}_{l+k,\nu} = 0, \qquad \nu=0,  \\
  & a\psi^{(1)}_{l+k,\nu} = [\nu(\nu+l+k+\tfrac{1}{2})]^{1/2} \psi^{(1)}_{l+k,\nu-1}, \qquad \nu=1, 2, \ldots.  \label{eq:action-RHO}
\end{split}
\end{equation}
\par
%
%
With the choice $(\varphi_1, \varphi_2, \ldots, \varphi_n) \to \bigl(\phi^{(l+k)}_{m_1}, \phi^{(l+k)}_{m_2}, \ldots, \phi^{(l+k)}_{m_k}\bigr)$, the partner potential (\ref{eq:partner}) of (\ref{eq:RHO}) is nonsingular if $m_1 < m_2 < \cdots < m_k$ with $m_i$ even (resp.~odd) for $i$ odd (resp.~even) and if  in addition $\alpha + k > m_k$. By proceeding as in Subsec.~IIA, it can be rewritten as
\begin{equation}
  V^{(2)}(x) = V_l(x) - k - 2 \frac{d^2}{dx^2} \log \tilde{\cal W}\bigl(L^{(-\alpha-k)}_{m_1}(-z), L^{(-\alpha-k)}_{m_2}(-z), \ldots, 
  L^{(-\alpha-k)}_{m_k}(-z)\bigr),  \label{eq:partner-add-bis}
\end{equation}
where $\tilde{\cal W}\bigl(f_1(z), f_2(z), \ldots, f_k(z)\bigr)$ denotes the Wronskian of the functions $f_1(z), f_2(z), \ldots, f_k(z)$ with respect to $z$. Its bound-state energies and corresponding wavefunctions are given by
\begin{equation}
  E^{(2)}_{l,\nu} = 2\nu + \alpha + k + 1, \qquad \nu=-m_k-1, \ldots, -m_2-1, -m_1-1, 0, 1, 2, \ldots,  
  \label{eq:partner-energy-bis}
\end{equation}
and
\begin{equation}
\begin{split}
  & \psi^{(2)}_{l,\nu}(x) \propto \frac{{\cal W}\bigl(\phi^{(l+k)}_{m_1}, \phi^{(l+k)}_{m_2}, \ldots, \phi^{(l+k)}_{m_k},
        \psi^{(l+k)}_{\nu}\bigr)}{{\cal W}\bigl(\phi^{(l+k)}_{m_1}, \phi^{(l+k)}_{m_2}, \ldots, \phi^{(l+k)}_{m_k}\bigr)}, \qquad
        \nu=0, 1, 2, \ldots, \\
  & \psi^{(2)}_{l,-m_i-1}(x) \propto \frac{{\cal W}\bigl(\phi^{(l+k)}_{m_1}, \phi^{(l+k)}_{m_2}, \ldots, \check{\phi}^{(l+k)}_{m_i},
        \ldots, \phi^{(l+k)}_{m_k}\bigr)}{{\cal W}\bigl(\phi^{(l+k)}_{m_1}, \phi^{(l+k)}_{m_2}, \ldots, \phi^{(l+k)}_{m_k}\bigr)}, 
        \qquad i=1, 2, \ldots, k,      
\end{split}
\end{equation}
respectively.\par
%
%
To obtain an (essentially) equivalent result for the partner potential derived in the state-deleting approach, we have to start this time from a potential $\bar{V}^{(1)}(x)$ that differs from $V^{(1)}(x)$, considered in (\ref{eq:RHO}), namely
\begin{equation}
  \bar{V}^{(1)}(x) = V_{l+k-m_k-1}(x),  \label{eq:V-1-bis}
\end{equation}
where we assume $\alpha + k > m_k + 1$. Then, with the choice $n = m_k + 1 - k$ and \linebreak 
$(\varphi_1, \varphi_2, \ldots, \varphi_n) \to \bigl(\psi^{(l+k-m_k-1)}_1, \psi^{(l+k-m_k-1)}_2, \ldots, \check{\psi}^{(l+k-m_k-1)}_{m_k-m_{k-1}}, \ldots,
\check{\psi}^{(l+k-m_k-1)}_{m_k-m_2}, \ldots,$ \linebreak 
$\check{\psi}^{(l+k-m_k-1)}_{m_k-m_1}, \ldots, \psi^{(l+k-m_k-1)}_{m_k}\bigr)$ in (\ref{eq:partner}), we get
\begin{equation}
\begin{split}
  \bar{V}^{(2)}(x) &= V_l(x) + m_k + 1 - k - 2 \frac{d^2}{dx^2} \log \tilde{{\cal W}}\bigl(L^{(\alpha+k-m_k-1)}_1(z),
     L^{(\alpha+k-m_k-1)}_2(z), \ldots, \\
  & \qquad \check{L}^{(\alpha+k-m_k-1)}_{m_k-m_{k-1}}(z), \ldots,  \check{L}^{(\alpha+k-m_k-1)}_{m_k-m_{1}}(z), \ldots,
     L^{(\alpha+k-m_k-1)}_{m_k}(z)\bigr), 
\end{split}  \label{eq:V-2-bis}
\end{equation}
which can be shown \cite{odake13a} to satisfy the relation
\begin{equation}
  V^{(2)}(x) + m_k + 1 = \bar{V}^{(2)}(x)  \label{eq:V-bar-2-bis}
\end{equation}
with the potential $V^{(2)}(x)$ given in (\ref{eq:partner-add-bis}). In (\ref{eq:V-bar-2-bis}), nonsingular potentials correspond to $\alpha + k > m_k + 1$ and $m_1 < m_2 < \cdots < m_k$ with $m_i$ even (resp.\ odd) for $i$ odd (resp.\ even). In the case of (\ref{eq:V-1-bis}) and (\ref{eq:V-2-bis}),  the bound-state energies and wavefunctions can be written as
\begin{equation}
\begin{split}
  & \bar{E}^{(1)}_{l+k-m_k-1,\nu} = 2\nu + \alpha + k - m_k, \qquad \nu=0, 1, 2, \ldots, \\
  & \bar{E}^{(2)}_{l,\nu} = 2\nu + \alpha + k + m_k + 2, \qquad \nu=-m_k-1, \ldots, -m_2-1, -m_1-1, 0, 1, 2, \ldots,
\end{split}  \label{eq:E-bar}
\end{equation}
and $\bar{\psi}^{(1)}_{l+k-m_k-1,\nu}(x) \propto \psi^{(l+k-m_k-1)}_{\nu}(x)$, as well as $\bar{\psi}^{(2)}_{l,\nu}(x)$ obtained from Eq.~(\ref{eq:partner-wf}) by changing $\psi_j(x)$ into $\psi^{(l+k-m_k-1)}_j(x)$.\par
%
%
As above-mentioned, an important difference between the linear and radial oscillator cases is that in the latter $\bar{H}^{(1)}$ and $H^{(1)}$ do not coincide. As for the construction of ladder operators in Sec.~III, it will be important to be able to relate these Hamiltonians, we show in Appendix A that this can be done, up to some energy shift, by a $(m_k+1)$th-order  SUSYQM transformation using seed functions of type II.\par
%
%
\section{LADDER OPERATORS FOR TYPE III MULTI-STEP RATIONALLY-EXTENDED POTENTIALS}

\setcounter{equation}{0}

In the present section, we will avail ourselves of the existence of two different paths going from $H^{(1)}$ to $H^{(2)}$ (up to some addition constant) to build some ladder operators for $H^{(2)}$.\par
%
%
\subsection{Harmonic oscillator case}

{}From Eqs.~(\ref{eq:V-bar-1}), (\ref{eq:V-bar-2}), and the intertwining relations (\ref{eq:intertwine}) (as well as their counterparts for $\bar{H}^{(1)}$, $\bar{H}^{(2)}$, $\bar{\cal A}$, and $\bar{\cal A}^{\dagger}$), it is clear that one can go from $H^{(2)}$ to $H^{(2)} + 2m_k + 2$ along the following path
\begin{equation}
  \xymatrixcolsep{5pc}\xymatrix@1{
  H^{(2)}  \ar@/_{10mm}/[rr]^{c}  \ar[r]^{{\cal A}^{\dagger}} & H^{(1)} = \bar{H}^{(1)}  \ar[r]^-{\bar{\cal A}} & \bar{H}^{(2)} =
  H^{(2)}+2m_k+2}  \label{eq:path}
\end{equation}
The $(m_k+1)$th-order differential operator
\begin{equation}
  c = \bar{\cal A} {\cal A}^{\dagger}  \label{eq:ladder-k-HO-1}
\end{equation}
that performs such a transformation is therefore a lowering operator for $H^{(2)}$. Together with its Hermitian conjugate, the creation operator
\begin{equation}
  c^{\dagger} = {\cal A} \bar{\cal A}^{\dagger}, \label{eq:ladder-k-HO-2}
\end{equation}
and $H^{(2)}$, it satisfies the commutation relations
\begin{equation}
\begin{split}
  & [H^{(2)}, c^{\dagger}] = (2m_k+2) c^{\dagger}, \qquad [H^{(2)}, c] = - (2m_k+2) c, \\
  & [c, c^{\dagger}] = Q(H^{(2)}+2m_k+2) - Q(H^{(2)}), 
\end{split}  \label{eq:PHA}
\end{equation}
defining a PHA of $m_k$th order \cite{fernandez99, carballo}. In Eq.~(\ref{eq:PHA}), $Q(H^{(2)}) = c^{\dagger} c$ is indeed a $(m_k+1)$th-order polynomial in $H^{(2)}$, which we now plan to determine.\par
%
%
{}First, let us observe that since ${\cal A}^{\dagger}$ annihilates the $k$ wavefunctions $\psi^{(2)}_{-m_i-1}(x)$, $i=1$, 2, \ldots, $k$, whose energy is $E^{(2)}_{-m_i-1} = -2m_i - 1$, we can express the $2k$th-order differential operator ${\cal A}{\cal A}^{\dagger}$ as
\begin{equation}
  {\cal A} {\cal A}^{\dagger} = \prod_{i=1}^k (H^{(2)} + 2m_i + 1).  \label{eq:rel-1}
\end{equation}
From the intertwining relations (\ref{eq:intertwine}), we then get
\begin{equation}
  {\cal A}^{\dagger} {\cal A} = \prod_{i=1}^k (H^{(1)} + 2m_i + 1).
\end{equation}
\par
%
%
Second, let us note that $\bar{\cal A}$ annihilates the $m_k+1-k$ excited states $\psi^{(1)}_j(x)$, $j=1$, 2, \ldots, $m_k-m_{k-1}-1$, $m_k-m_{k-1}+1$, \ldots, $m_k-m_1-1$, $m_k-m_1+1$, \ldots, $m_k$, of $\bar{H}^{(1)} = H^{(1)}$, corresponding to the energies $E^{(1)}_j = 2j+1$; hence, the $2(m_k+1-k)$th-order differential operator $\bar{\cal A}^{\dagger} \bar{\cal A}$ can be written as
\begin{equation}
  \bar{\cal A}^{\dagger} \bar{\cal A} = \prod_{\substack{j=1 \\ j \ne m_k-m_{k-1}, \ldots, m_k-m_1}}^{m_k} 
  (\bar{H}^{(1)} - 2j -1), 
\end{equation}
with a similar expression for $\bar{\cal A} \bar{\cal A}^{\dagger}$ in terms of $\bar{H}^{(2)}$. In terms of $H^{(1)}$ and $H^{(2)}$, this leads to the relations
\begin{equation}
  \bar{\cal A}^{\dagger} \bar{\cal A} = \prod_{\substack{j=1 \\ j \ne m_k-m_{k-1}, \ldots, m_k-m_1}}^{m_k} (H^{(1)} - 2j -1), 
\end{equation}
\begin{equation}
  \bar{\cal A} \bar{\cal A}^{\dagger} = \prod_{\substack{j=1 \\ j \ne m_k-m_{k-1}, \ldots, m_k-m_1}}^{m_k} (H^{(2)} 
  + 2m_k - 2j +1).  \label{eq:rel-2} 
\end{equation}
\par
%
%
It only remains to combine Eqs.~(\ref{eq:rel-1})--(\ref{eq:rel-2}) with the intertwining relations to obtain
\begin{equation}
  Q(H^{(2)}) = \left(\prod_{i=1}^k (H^{(2)} + 2m_i + 1)\right) \left(\prod_{\substack{j=1 \\ j \ne m_k-m_{k-1}, \ldots, m_k-m_1}}
  ^{m_k} (H^{(2)} - 2j -1)\right),  \label{eq:Q}
\end{equation}
which completes the definition of the PHA (\ref{eq:PHA}).\par
%
%
The action of the operator $Q(H^{(2)})$ on the wavefunctions $\psi^{(2)}_{\nu}(x)$, $\nu=-m_k-1$, \ldots, $-m_1-1$, 0, 1, 2, \ldots, is readily obtained by replacing $H^{(2)}$ on the right-hand side of (\ref{eq:Q}) by the corresponding eigenvalues $E^{(2)}_{\nu}$, given in Eq.~(\ref{eq:energy-add}). If we choose the normalization constants of the $\psi^{(2)}_{\nu}$'s in such a way that all matrix elements of the ladder operators $c^{\dagger}$, $c$ are nonnegative, then such an action yields the following results:
\begin{equation}
\begin{split}
  & c\psi^{(2)}_{\nu} = 0, \qquad \nu=-m_k-1, \ldots, -m_1-1, 1, 2, \ldots, m_k-m_{k-1}-1,  \\
  & \qquad  m_k-m_{k-1}+1, \ldots, m_k-m_1-1, m_k-m_1+1, \ldots, m_k,  \label{eq:action-0}
\end{split}
\end{equation}
\begin{equation}
  c\psi^{(2)}_0 = \Biggl[2^{m_k+1} (m_k+1)! \Biggl(\prod_{i=1}^{k-1} \frac{m_i+1}{m_k-m_i}\Biggr)\Biggr]^{1/2} 
  \psi^{(2)}_{-m_k-1},  \label{eq:action-1} 
\end{equation}
\begin{equation}
\begin{split}
  & c\psi^{(2)}_{m_k-m_i} =  \Biggl[2^{m_k+1} (m_k+1) (2m_k-m_i+1) (m_k-m_i-1)! m_i! \\
  & \quad \times \Biggl(\prod_{j=1}^{i-1} \frac{m_k+m_j-m_i+1}{m_i-m_j}\Biggr) \Biggl(\prod_{l=i+1}^{k-1} 
       \frac{m_k+m_l-m_i+1}{m_l-m_i}\Biggr)\Biggr]^{1/2} \psi^{(2)}_{-m_i-1}, \\
  & \quad \qquad i=1, 2, \ldots, k-1, 
\end{split}
\end{equation}
\begin{equation}
\begin{split}
  & c\psi^{(2)}_{\nu} =  \Biggl[2^{m_k+1} (\nu+m_k+1) \frac{(\nu-1)!}{(\nu-m_k-1)!} \Biggl(\prod_{i=1}^{k-1} \frac{\nu+m_i+1}
        {\nu+m_i-m_k}\Biggr)\Biggr]^{1/2} \psi^{(2)}_{\nu-m_k-1}, \\
  & \quad \qquad \nu=m_k+1, m_k+2, \ldots.
\end{split}  \label{eq:action-2}
\end{equation}   
The action of $c^{\dagger}$ on $\psi^{(2)}_{\nu}(x)$ is finally deduced from Eqs.~(\ref{eq:action-1})--(\ref{eq:action-2}) by using Hermitian conjugation.\par
%
%
We conclude that the PHA generated by $H^{(2)}$, $c^{\dagger}$, and $c$ has $m_k+1$ infinite-dimensional unirreps spanned by the states $\{\psi^{(2)}_{i+(m_k+1) j} \mid j=0,1,2,\ldots\}$ with $i=-m_k-1$, \ldots, $-m_1-1$, 1, 2, \ldots, $m_k-m_{k-1}-1$, $m_k-m_{k-1}+1$, \ldots, $m_k-m_1-1$, $m_k-m_1+1$, \ldots, $m_k$, respectively.\par
%
%
It is worth observing that the ladder operators $c^{\dagger}$, $c$, constructed in this subsection, extend to arbitrary $k$ values those proposed in Ref.~\cite{marquette13c} for $k=1$.\par
%
%
\subsection{Radial harmonic oscillator case}

The construction of ladder operators for the Hamiltonian $H^{(2)}$, obtained from Eq.~(\ref{eq:partner-add-bis}), is similar to that performed in Subsec.~IIIA, except for the need of using an intermediate transformation from $H^{(1)}$ to $\bar{H}^{(1)}$ by means of the additional intertwining operators $\tilde{\cal A}$ and $\tilde{\cal A}^{\dagger}$, derived in Appendix A. Instead of (\ref{eq:path}), let us therefore consider the following path
\begin{equation}
  \xymatrixcolsep{1.5pc}\xymatrix@1{
  H^{(2)}  \ar@/_{10mm}/[rrr]^{c}  \ar[r]^-{{\cal A}^{\dagger}} & H^{(1)}=\tilde{H}^{(1)}  \ar[r]^-{\tilde{\cal A}}  & \tilde{H}^{(2)} =
  \bar{H}^{(1)}+m_{k}+1 \ar[r]^-{\bar{\cal A}}  & \bar{H}^{(2)}+m_{k}+1=H^{(2)}+2m_{k}+2}
\end{equation}
The corresponding lowering operator $c$ and its Hermitian conjugate $c^{\dagger}$ are now defined by
\begin{equation}
  c = \bar{\cal A} \tilde{\cal A} {\cal A}^{\dagger}, \qquad c^{\dagger} = {\cal A} \tilde{\cal A}^{\dagger} \bar{\cal A}^{\dagger}, \label{eq:ladder-k-RHO-1}
\end{equation}
respectively. They are $(2m_k+2)$th-order differential operators, closing a PHA of $(2m_k+1)$th order with $H^{(2)}$. For such a PHA, Eq.~(\ref{eq:PHA}) remains valid with an expression of $Q(H^{(2)}) = c^{\dagger} c$ different from (\ref{eq:Q}).\par
%
%
To derive the latter, let us observe that an argument similar to that used in Subsec.~IIIA leads to the equations
\begin{equation}
\begin{split}
  {\cal A} {\cal A}^{\dagger} &= \prod_{i=1}^k (H^{(2)} - \alpha - k + 2m_i + 1), \\
  {\cal A}^{\dagger} {\cal A} &= \prod_{i=1}^k (H^{(1)} - \alpha - k + 2m_i + 1) = \prod_{i=1}^k (\tilde{H}^{(1)} - \alpha - k + 2m_i 
       + 1), \\
  \bar{\cal A}^{\dagger} \bar{\cal A} &= \prod_{\substack{n=1 \\ n \ne m_k-m_{k-1}, \ldots, m_k-m_1}}^{m_k} 
       (\bar{H}^{(1)} - \alpha - 2n + m_k - k) \\
  &= \prod_{\substack{n=1 \\ n \ne m_k-m_{k-1}, \ldots, m_k-m_1}}^{m_k} (\tilde{H}^{(2)} - \alpha - 2n - k - 1), \\
  \bar{\cal A} \bar{\cal A}^{\dagger} &= \prod_{\substack{n=1 \\ n \ne m_k-m_{k-1}, \ldots, m_k-m_1}}^{m_k} 
       (\bar{H}^{(2)} - \alpha - 2n + m_k - k) \\
  &= \prod_{\substack{n=1 \\ n \ne m_k-m_{k-1}, \ldots, m_k-m_1}}^{m_k} (H^{(2)} - \alpha - 2n + 2m_k - k + 1).     
\end{split}  \label{eq:rel-3}
\end{equation}
In the same way, since the operator $\tilde{\cal A}$ annihilates the nonnormalizable states $\tilde{\phi}^{(l+k)}_j(x)$, $j=0$, 1, \ldots, $m_k$, of energy $E^{\rm II}_{l+k,j} = - \alpha - k + 2j + 1$, considered in Appendix A, we can express the $(2m_k+2)$th-order differential operator $\tilde{\cal A}^{\dagger} \tilde{\cal A}$ as
\begin{equation}
  \tilde{\cal A}^{\dagger} \tilde{\cal A} = \prod_{j=0}^{m_k} (\tilde{H}^{(1)} + \alpha + k - 2j - 1) = \prod_{j=0}^{m_k} (H^{(1)} 
  + \alpha + k - 2j - 1). 
\end{equation}
From the intertwining relations, we then also get
\begin{equation}
  \tilde{\cal A} \tilde{\cal A}^{\dagger} = \prod_{j=0}^{m_k} (\tilde{H}^{(2)} + \alpha + k - 2j - 1) = \prod_{j=0}^{m_k} 
  (\bar{H}^{(1)} + \alpha + k  + m_k - 2j).  \label{eq:rel-4} 
\end{equation}
\par
%
%
Equations (\ref{eq:rel-3})--(\ref{eq:rel-4}) finally yield
\begin{equation}
\begin{split}
  Q(H^{(2)}) &= \left(\prod_{i=1}^k (H^{(2)} - \alpha + 2m_i - k + 1)\right) \left(\prod_{j=0}^{m_k} (H^{(2)} + \alpha - 2j + k 
       - 1)\right) \\
  & \quad \times \left(\prod_{\substack{n=1 \\ n \ne m_k-m_{k-1}, \ldots, m_k-m_1}}^{m_k} (H^{(2)} - \alpha - 2n - k -1)\right).  
\end{split}  \label{eq:Q-bis}
\end{equation}  
Let us stress that this relation and Eq.~(\ref{eq:PHA}) have to be completed with the condition $\alpha + k > m_k + 1$.\par
%
%
On proceeding as in Subsec.~IIIA, from Eqs.~(\ref{eq:partner-energy-bis}) and (\ref{eq:Q-bis}), we can easily derive the following results for the action of $c$ on the wavefunctions $\psi^{(2)}_{l,\nu}(x)$, $\nu=-m_k-1$, \ldots, $-m_1-1$, 0, 1, 2, \ldots, of $H^{(2)}$:
\begin{equation}
\begin{split}
  & c\psi^{(2)}_{l,\nu} = 0, \qquad \nu=-m_k-1, \ldots, -m_1-1, 1, 2, \ldots, m_k-m_{k-1}-1,  \\
  & \qquad  m_k-m_{k-1}+1, \ldots, m_k-m_1-1, m_k-m_1+1, \ldots, m_k,  
\end{split} \label{eq:action-0-bis}
\end{equation}
\begin{equation}
  c\psi^{(2)}_{l,0} = 2^{m_k+1} \Biggl[(m_k+1)! \frac{\Gamma(\alpha+k+1)}{\Gamma(\alpha+k-m_k)} \Biggl(\prod_{i=1}^{k-1}
  \frac{m_i+1}{m_k-m_i}\Biggr)\Biggr]^{1/2} \psi^{(2)}_{l,-m_k-1},   
\end{equation}
\begin{equation}
\begin{split}
  & c\psi^{(2)}_{l,m_k-m_i} =  2^{m_k+1} \Biggl[(m_k+1) (2m_k-m_i+1) (m_k-m_i-1)! m_i! \\
  & \quad \times  \frac{\Gamma(\alpha+m_k-m_i+k+1)}{\Gamma(\alpha-m_i+k)} 
       \Biggl(\prod_{j=1}^{i-1} \frac{m_k+m_j-m_i+1}{m_i-m_j}\Biggr) \\
  & \quad \times \Biggl(\prod_{n=i+1}^{k-1} \frac{m_k+m_n-m_i+1}{m_n-m_i}\Biggr)\Biggr]^{1/2} \psi^{(2)}_{l,-m_i-1}, \qquad
       i=1, 2, \ldots, k-1, 
\end{split}
\end{equation}
\begin{equation}
\begin{split}
  & c\psi^{(2)}_{l,\nu} =  2^{m_k+1} \Biggl[(\nu+m_k+1) \frac{(\nu-1)!}{(\nu-m_k-1)!} \frac{\Gamma(\nu+\alpha+k+1)}
        {\Gamma(\nu+\alpha+k-m_k)} \\ 
  & \quad \times \Biggl(\prod_{i=1}^{k-1} \frac{\nu+m_i+1}{\nu+m_i-m_k}\Biggr)\Biggr]^{1/2} \psi^{(2)}_{l,\nu-m_k-1}, \qquad 
        \nu=m_k+1, m_k+2, \ldots,
\end{split} \label{eq:action-2-bis} 
\end{equation}
and corresponding results for the action of $c^{\dagger}$.\par
%
%
When we compare Eqs.~(\ref{eq:action-0-bis})--(\ref{eq:action-2-bis}) with the corresponding results for the extended harmonic oscillator, contained in Eqs.~(\ref{eq:action-0})--(\ref{eq:action-2}), we note that there are very few changes, namely all matrix elements of $c$ given in (\ref{eq:action-1})--(\ref{eq:action-2}) are multiplied by a generic factor $[2^{m_k+1} \Gamma(\nu+\alpha+k+1)/\Gamma(\nu+\alpha +k-m_k)]^{1/2}$, which is nonvanishing for any $\nu \in \{0, m_k-m_{k-1}, \ldots, m_k-m_1, m_k+1, m_k+2, \ldots\}$. Hence, as before, the present PHA has $m_k+1$ infinite-dimensional unirreps, spanned this time by the states $\{\psi^{(2)}_{l,i+(m_k+1) j} \mid j=0,1,2,\ldots\}$ with $i=-m_k-1$, \ldots, $-m_1-1$, 1, 2, \ldots, $m_k-m_{k-1}-1$, $m_k-m_{k-1}+1$, \ldots, $m_k-m_1-1$, $m_k-m_1+1$, \ldots, $m_k$, respectively.\par
%
%
The energy levels of the three pairs of partner Hamiltonians $(H^{(1)}, H^{(2)})$, $(\tilde{H}^{(1)}, \tilde{H}^{(2)})$, and $(\bar{H}^{(1)}, \bar{H}^{(2)})$, as well as the action of the corresponding intertwining operators, are displayed in Fig.~1 for the case where $k=2$, $m_1=2$, $m_2=3$, $l=2$, and $\alpha=5/2$. The resulting four PHA unirreps are shown in Fig.~2.\par
%
%
\begin{center}
\begin{picture}(160,130)(0,0)

\put(10,5){\thicklines\vector(0,1){120}}
\multiput(10,5)(0,24){5}{\ord}
\put(5,5){\makebox(4,0)[r]{\large -5/2}}
\put(5,29){\makebox(4,0)[r]{\large 11/2}}
\put(5,53){\makebox(4,0)[r]{\large 27/2}}
\put(5,77){\makebox(4,0)[r]{\large 43/2}}
\put(5,101){\makebox(4,0)[r]{\large 59/2}}
\put(5,120){\makebox(4,0)[r]{\large $E$}}

\put(20,5){\niveau}
\put(20,11){\niveau}
\multiput(20,29)(0,6){13}{\niveau}
\put(37,5){\makebox(4,0)[r]{\large -4}}
\put(37,11){\makebox(4,0)[r]{\large -3}}
\put(37,29){\makebox(4,0)[r]{\large 0}}
\put(37,35){\makebox(4,0)[r]{\large 1}}
\put(37,41){\makebox(4,0)[r]{\large 2}}
\put(37,47){\makebox(4,0)[r]{\large 3}}
\put(37,53){\makebox(4,0)[r]{\large 4}}
\put(37,59){\makebox(4,0)[r]{\large 5}}
\put(37,65){\makebox(4,0)[r]{\large 6}}
\put(37,71){\makebox(4,0)[r]{\large 7}}
\put(37,77){\makebox(4,0)[r]{\large 8}}
\put(37,83){\makebox(4,0)[r]{\large 9}}
\put(37,89){\makebox(4,0)[r]{\large 10}}
\put(37,95){\makebox(4,0)[r]{\large 11}}
\put(37,101){\makebox(4,0)[r]{\large 12}}
\put(28,0){\makebox(4,0)[r]{$H^{(2)}$}}

\multiput(60,29)(0,6){13}{\niveau}
\put(77,29){\makebox(4,0)[r]{\large 0}}
\put(77,35){\makebox(4,0)[r]{\large 1}}
\put(77,41){\makebox(4,0)[r]{\large 2}}
\put(77,47){\makebox(4,0)[r]{\large 3}}
\put(77,53){\makebox(4,0)[r]{\large 4}}
\put(77,59){\makebox(4,0)[r]{\large 5}}
\put(77,65){\makebox(4,0)[r]{\large 6}}
\put(77,71){\makebox(4,0)[r]{\large 7}}
\put(77,77){\makebox(4,0)[r]{\large 8}}
\put(77,83){\makebox(4,0)[r]{\large 9}}
\put(77,89){\makebox(4,0)[r]{\large 10}}
\put(77,95){\makebox(4,0)[r]{\large 11}}
\put(77,101){\makebox(4,0)[r]{\large 12}}
\put(76,0){\makebox(4,0)[r]{$H^{(1)} = \tilde{H}^{(1)}$}}

\multiput(100,29)(0,6){13}{\niveau}
\put(117,29){\makebox(4,0)[r]{\large 0}}
\put(117,35){\makebox(4,0)[r]{\large 1}}
\put(117,41){\makebox(4,0)[r]{\large 2}}
\put(117,47){\makebox(4,0)[r]{\large 3}}
\put(117,53){\makebox(4,0)[r]{\large 4}}
\put(117,59){\makebox(4,0)[r]{\large 5}}
\put(117,65){\makebox(4,0)[r]{\large 6}}
\put(117,71){\makebox(4,0)[r]{\large 7}}
\put(117,77){\makebox(4,0)[r]{\large 8}}
\put(117,83){\makebox(4,0)[r]{\large 9}}
\put(117,89){\makebox(4,0)[r]{\large 10}}
\put(117,95){\makebox(4,0)[r]{\large 11}}
\put(117,101){\makebox(4,0)[r]{\large 12}}
\put(117,0){\makebox(4,0)[r]{$\tilde{H}^{(2)} = \bar{H}^{(1)} + 4$}}

\put(140,29){\niveau}
\put(140,35){\niveau}
\multiput(140,53)(0,6){9}{\niveau}
\put(157,29){\makebox(4,0)[r]{\large -4}}
\put(157,35){\makebox(4,0)[r]{\large -3}}
\put(157,53){\makebox(4,0)[r]{\large 0}}
\put(157,59){\makebox(4,0)[r]{\large 1}}
\put(157,65){\makebox(4,0)[r]{\large 2}}
\put(157,71){\makebox(4,0)[r]{\large 3}}
\put(157,77){\makebox(4,0)[r]{\large 4}}
\put(157,83){\makebox(4,0)[r]{\large 5}}
\put(157,89){\makebox(4,0)[r]{\large 6}}
\put(157,95){\makebox(4,0)[r]{\large 7}}
\put(157,101){\makebox(4,0)[r]{\large 8}}
\put(160,0){\makebox(4,0)[r]{$\bar{H}^{(2)} + 4 = H^{(2)} + 8$}}

\put(27.5,107){\usebox{\dash}}
\put(67.5,107){\usebox{\dash}}
\put(107.5,107){\usebox{\dash}}
\put(147.5,107){\usebox{\dash}}

\multiput(45,29)(40,0){3}{\arrow}
\put(48,33){\makebox(4,0)[r]{\large ${\cal A}^{\dagger}$}}
\put(87,33){\makebox(4,0)[r]{\large $\tilde{{\cal A}}$}}
\put(127,33){\makebox(4,0)[r]{\large $\bar{{\cal A}}$}}

\end{picture}

\bigskip
\noindent

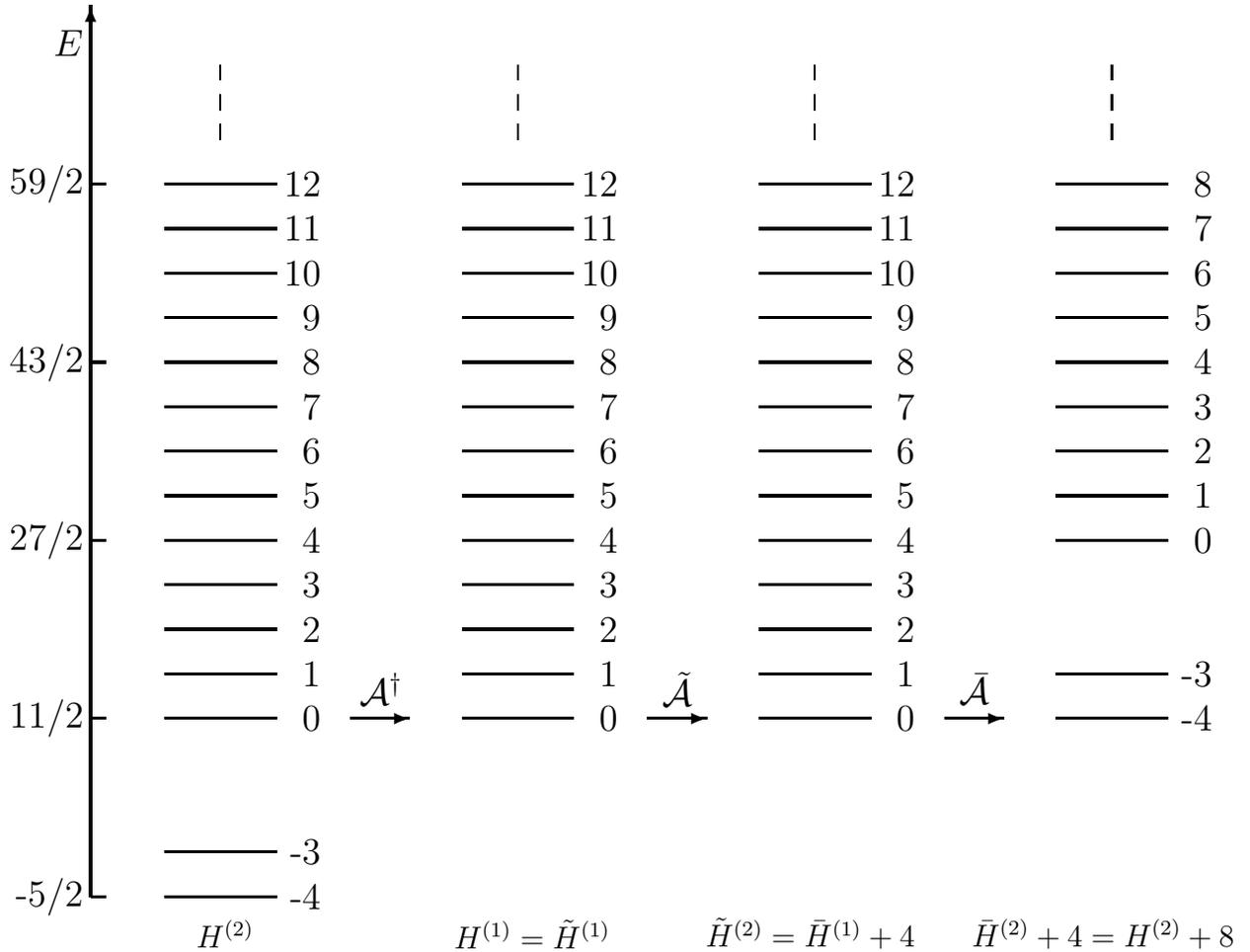
Fig.~1. Energy spectra of $H^{(2)}$, $H^{(1)}$, $\tilde{H}^{(2)}$, $H^{(2)} + 8$, and action of ${\cal A}^{\dagger}$, $\tilde{\cal A}$, $\bar{\cal A}$ for $m_1=2$, $m_2=3$, $\alpha=5/2$. The $\nu$ values are indicated on the right.

\end{center}

\par
%
%
\begin{center}
\begin{picture}(100,100)(0,0)

\put(10,5){\thicklines\vector(0,1){120}}
\multiput(10,5)(0,24){5}{\ord}
\put(5,5){\makebox(4,0)[r]{\large -5/2}}
\put(5,29){\makebox(4,0)[r]{\large 11/2}}
\put(5,53){\makebox(4,0)[r]{\large 27/2}}
\put(5,77){\makebox(4,0)[r]{\large 43/2}}
\put(5,101){\makebox(4,0)[r]{\large 59/2}}
\put(5,120){\makebox(4,0)[r]{\large $E^{(2)}_{2,\nu}$}}

\put(20,5){\level}
\put(20,11){\level}
\multiput(20,29)(0,6){13}{\level}
\put(30,107){\usebox{\dash}}
\put(50,107){\usebox{\dash}}
\put(70,107){\usebox{\dash}}
\put(90,107){\usebox{\dash}}

\multiput(30,5)(0,24){4}{\fleche}
\multiput(50,11)(0,24){3}{\fleche}
\put(50,83){\thicklines\line(0,1){18}}
\multiput(70,41)(0,24){2}{\fleche}
\put(70,89){\thicklines\line(0,1){12}}
\multiput(90,47)(0,24){2}{\fleche}
\put(90,95){\thicklines\line(0,1){6}}

\put(104,5){\makebox(4,0)[r]{\large -4}}
\put(104,11){\makebox(4,0)[r]{\large -3}}
\put(104,29){\makebox(4,0)[r]{\large 0}}
\put(104,35){\makebox(4,0)[r]{\large 1}}
\put(104,41){\makebox(4,0)[r]{\large 2}}
\put(104,47){\makebox(4,0)[r]{\large 3}}
\put(104,53){\makebox(4,0)[r]{\large 4}}
\put(104,59){\makebox(4,0)[r]{\large 5}}
\put(104,65){\makebox(4,0)[r]{\large 6}}
\put(104,71){\makebox(4,0)[r]{\large 7}}
\put(104,77){\makebox(4,0)[r]{\large 8}}
\put(104,83){\makebox(4,0)[r]{\large 9}}
\put(104,89){\makebox(4,0)[r]{\large 10}}
\put(104,95){\makebox(4,0)[r]{\large 11}}
\put(104,101){\makebox(4,0)[r]{\large 12}}

\end{picture}

\bigskip
\noindent

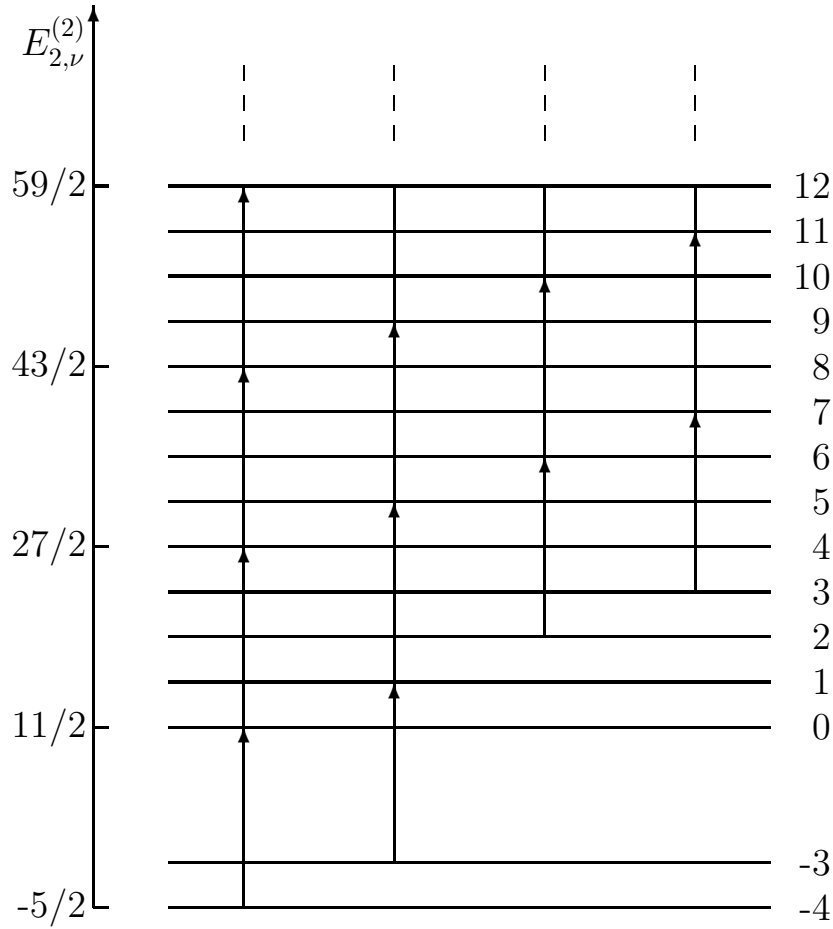
Fig.~2. Action of $c^{\dagger}$ on the $H^{(2)}$ eigenstates for $m_1=2$, $m_2=3$, $\alpha=5/2$. The $\nu$ values are indicated on the right.

\end{center}

\par
%
%
\section{APPLICATION TO SOME SUPERINTEGRABLE TWO-DIMENSIONAL SYSTEMS}

\setcounter{equation}{0}
 
\subsection{Superintegrability and polynomial algebras}

It is known from previous papers that integrals of motion of two-dimensional Hamiltonians in cartesian \cite{marquette10} or polar \cite{kalnins11} coordinates can be constructed using ladder operators or recurrence
relations for one-dimensional Hamiltonians. Here, we restrict the treatment to the case of two-dimensional Hamiltonians allowing separation of variables in cartesian
coordinates with scalar potentials of the form
\begin{equation}
  H = H_x + H_y = - \frac{d^2}{dx^2} - \frac{d^2}{dy^2} + V_x(x) + V_y(y),  \label{eq:H}
\end{equation}
and we assume that there exist ladder operators $(a_x^{\dagger}, a_x)$ and $(a_y^{\dagger}, a_y)$ in both axes that are differential operators of order $k_1$ and $k_2$, respectively, and satisfy the defining relations of two PHA's, \par
\begin{equation}
\begin{split}
  & [H_x, a_x^{\dagger}] = \lambda_x a_x^{\dagger}, \qquad [H_x, a_x] = - \lambda_x a_x, \qquad
      [a_x, a_x^{\dagger}] = Q(H_x + \lambda_x) - Q(H_x), \\
  & [H_y, a_y^{\dagger}] = \lambda_y a_y^{\dagger}, \qquad [H_y, a_y] = - \lambda_y a_y, \qquad
      [a_y, a_y^{\dagger}] = S(H_y + \lambda_y) - S(H_y).      
\end{split}  \label{eq:PHA-PHA}
\end{equation}
In Eq.~(\ref{eq:PHA-PHA}), the constants $\lambda_x$ and $\lambda_y$ correspond to the shift in the energy of the wavefunctions by  application of the ladder operators. By construction,
such ladder operators only admit combinations of finite or infinite sequences of equidistant levels. However, interesting patterns can be created by these combinations. In Eq.~(\ref{eq:PHA-PHA}), $Q(H_x)$ and $S(H_y)$ are polynomials that can be expressed further as 
\begin{equation}
  Q(H_x)=\prod_{i=1}^{k_{1}}(H_{x}-\epsilon_{i}^{(x)}),  \label{eq:Qx}
\end{equation}
\begin{equation}
  S(H_y)=\prod_{j=1}^{k_{2}}(H_{y}-\epsilon_{j}^{(y)}).  \label{eq:Sy}
\end{equation}
The constants $\epsilon_{i}^{(x)}$ and $\epsilon_{j}^{(y)}$ will be related with zero modes of the lowering operators.
As seen in earlier sections, and also previous papers \cite{marquette13b,marquette13c}, ladder operators can themselves have an underlying structure. 
This can take the form of products of supercharges of various types. The intertwining and factorization relations satisfied by these supercharges can thus simplify the calculation of the polynomial Heisenberg algebra in both axes. \par
%
%
The separation of variables in cartesian coordinates implies the existence of a second-order integral of motion $H_x - H_y$, 
showing that the two-dimensional system (\ref{eq:H}) is integrable. From the ladder operators, one can construct additional polynomial operators commuting with
 $H$, $a_x^{\dagger n_1} a_y^{n_2}$ and $a_x^{n_1} a_y^{\dagger n_2}$, where $n_1$, $n_2 \in \mathbb{Z}^{+}$ are chosen such that 
$n_1 \lambda_x = n_2 \lambda_y = \bar{\lambda}$ \cite{marquette10}. Hence system (\ref{eq:H}) possesses three algebraically independent integrals of motion 
and thus has the superintegrability property. \par
%
%
The integrals of motion, written in the following factorized form
\begin{equation}
  K = \frac{1}{2\bar{\lambda}}(H_x-H_y),\qquad I_+ = a_x^{\dagger n_1} a_y^{n_2}, \qquad I_- = a_x^{n_1}
  a_y^{\dagger n_2},  \label{eq:integrals}
\end{equation}
generate the polynomial algebra of the system
\begin{equation}
\begin{split}
  & [K,I_{\pm}] = \pm I_{\pm}, \qquad [I_-,I_+] = F_{n_1,n_2}(K+1,H) - F_{n_1,n_2}(K,H), \\ 
  & F_{n_1,n_2}(K,H) = \prod_{i=1}^{n_1} Q\left(\frac{H}{2}+\bar{\lambda} K-(n_1-i)\lambda_{x}\right)
       \prod_{j=1}^{n_2}S\left(\frac{H}{2}-\bar{\lambda} K+j\lambda_y\right),
\end{split} \label{eq:algebra}
\end{equation}
which is of order $k_1n_1+k_2n_2-1$. This is the symmetry algebra of the two-dimensional superintegrable systems (\ref{eq:H}).\par
%
%
\subsection{Representations of the polynomial algebras}

In previous papers \cite{marquette13b,marquette13c,marquette09a,marquette09b,marquette10}, the Daskaloyannis approach \cite{daska01} was used in order to 
obtain finite-dimensional unirreps of polynomial algebras for various 
superintegrable Hamiltonians and furthermore to calculate algebraically the energy spectrum and the level total degeneracies. 
This approach is based on calculating the Casimir operator, establishing realizations of the polynomial algebras as deformed oscillator ones, generated by $\{N,b,b^{\dagger},1\}$ with commutation relations
\begin{equation}
  [N,b^{\dagger}]=b^{\dagger} ,\qquad [N,b]=-b ,\qquad b^{\dagger}b=\Phi(N,E,u) , \qquad bb^{\dagger}=\Phi(N+1,E,u),
\end{equation}
and imposing three constraints on the structure function $\Phi(N,E,u)$ (where $N$ is the number operator). These constraints are
\begin{equation}
  \Phi(0,E,u)=0,\qquad \Phi(p+1,E,u)=0, \qquad \Phi(x,E,u)>0,\qquad x=1,\ldots,p.
\end{equation}
\par
%
%
The resulting system of equations formed by these constraints involves the representation-dependent constant $u$, the energy $E$, the integer $p$, and can be solved, but we need to remove some 
non-physical solutions. In addition, it is not guaranteed the whole spectrum can be recovered, as in various cases the method does not provide the entire spectrum
and degeneracies \cite{marquette13b,marquette09a,marquette09b,marquette10}.\par
%
%
In a recent paper \cite{marquette13c}, we demonstrated that this method can be applied to families of systems related with one-step extensions of the harmonic
oscillator, containing as special cases Gravel systems for which no algebraic derivation of the complete energy spectrum and degeneracies was previously available \cite{gravel04,marquette09a,marquette09b}. We showed that such an algebraic
calculation can be carried out, provided a new set of integrals based on ladder operators with a different structure in terms of zero modes was constructed. We also obtained that unions of unirreps needed to be considered
to obtain the total degeneracies, a phenomenon not observed in quadratically superintegrable systems. These results also showed how the Daskaloyannis method can have wider applications
to systems with higher-order integrals. A negative point of this approach, however, remained the existence of nonphysical solutions that had to be eliminated. \par
%
%
The Daskaloyannis technique is particularly convenient in the case no underlying structure of integrals of motion is known. However, as described in Subsec.~IVA, we consider here some systems (\ref{eq:H}) whose individual constituents $H_x$ and $H_y$ have known ladder operators. The integrals of motion are thus products of the latter, as shown in Eq.~(\ref{eq:integrals}).
In addition, the ladder operators used are themselves composed of supercharges, so that their action on physical states is known (see Eqs.~(\ref{eq:action-0})--(\ref{eq:action-2}) and (\ref{eq:action-0-bis})--(\ref{eq:action-2-bis})). Taking advantage of this knowledge will therefore provide us with an alternative method to study the polynomial algebra representations. \par
%
%
The finite-dimensional unirreps of the polynomial algebra given by (\ref{eq:algebra}) (which is a deformed $u(2)$) may be characterized by $(N,s)$ and their basis states by
\begin{equation}
  |N, \tau,s, \sigma \rangle, \label{eq:basisunirreps}
\end{equation}
where $\sigma=-s,-s+1,\ldots, s$ and $\tau$ distinguishes between repeated representations specified by the same $s$. Here $s$ may be any integer or half-integer, $\sigma$ denotes the eigenvalue of 
$I_{0}=K+c$, where $c$ is some representation-dependent constant, and $I_{+}$, $I_{-}$ are such that
\begin{equation}
  I_{+}|N, \tau,s,s \rangle=I_{-}|N, \tau,s,-s \rangle =0. \label{eq:rep}
\end{equation}
From the last properties, one can find the value of $s$ realized for every $N$ value.  \par
%
%
In order to implement the method, we will denote the eigenstates of $H$ as $|N, \nu_{x}  \rangle$, where $N=\nu_{x}+\nu_{y}+1$ and $\nu_x$, $\nu_y$ are two quantum numbers associated with the eigenvalues of $H_x$ and $H_y$, respectively, and whose allowed values will depend on the examples considered. Equation~(\ref{eq:rep}) will be solved by deriving the action of integrals from that of ladder operators on physical states in the $|N, \nu_{x}  \rangle$ basis. Starting from states annihilated by 
$I_{-}$ and acting iteratively with $I_{+}$ until we reach a state annihilated by the latter, we identify the corresponding value of $s$ and the $\sigma$ associated with each state forming this sequence. Moreover, on using the notation $N= \lambda n_{1}n_{2}+\mu$ with some appropriate values of $\lambda$ and $\mu$, the basis states $|N, \tau,s, \sigma \rangle$ are obtained (with values of $\tau$ related to values of $\lambda$ and $\mu$). In this way of deducing finite-dimensional unirreps, we do not need to remove nonphysical states as such a method only involves action on physical ones. \par
%
%
\subsection{\boldmath Combination of a $k$-step rationally-extended potential with a standard one}

Let us consider 2D Hamiltonians of the form given by Eq.~(\ref{eq:H}). In the $x$-axis, we take a $k$-step extension of 
the harmonic or radial oscillator, given by Eq.~(\ref{eq:partner-add}) or Eq.~(\ref{eq:partner-add-bis}), respectively. In the
$y$-axis, we simply take the harmonic or radial oscillator. This provides us with four new infinite families of 2D superintegrable systems:
\begin{equation}
  H_{a}= - \frac{d^2}{dx^2} - \frac{d^2}{dy^2} + x^2  + y^2 - 2k - 2 \frac{d^2}{dx^2} 
  \log {\cal W}({\cal H}_{m_1}, {\cal H}_{m_2}, \ldots, {\cal H}_{m_k}),
\end{equation}
\begin{equation}
\begin{split}
  H_{b} &=  - \frac{d^2}{dx^2} - \frac{d^2}{dy^2} + \frac{1}{4} x^2 + \frac{l(l+1)}{x^2} + y^2 - k \\
  & \quad - 2 \frac{d^2}{dx^2} \log \tilde{\cal W}\bigl(L^{(-\alpha-k)}_{m_1}(-z), L^{(-\alpha-k)}_{m_2}(-z), \ldots,
       L^{(-\alpha-k)}_{m_k}(-z)\bigr) ,
\end{split}
\end{equation}
\begin{equation}
  H_{c}= - \frac{d^2}{dx^2} - \frac{d^2}{dy^2}+ x^2 + \frac{1}{4} y^2 + \frac{l(l+1)}{y^2} - 2k 
  - 2 \frac{d^2}{dx^2} \log {\cal W}({\cal H}_{m_1}, {\cal H}_{m_2}, \ldots,{\cal H}_{m_k}),
\end{equation}
\begin{equation}
\begin{split}
  H_{d} &=  - \frac{d^2}{dx^2} - \frac{d^2}{dy^2}+ \frac{1}{4} x^2 + \frac{l(l+1)}{x^2} +  \frac{1}{4} y^2 + \frac{l(l+1)}{y^2} - k \\
  & \quad - 2 \frac{d^2}{dx^2} \log \tilde{\cal W}\bigl(L^{(-\alpha-k)}_{m_1}(-z), L^{(-\alpha-k)}_{m_2}(-z), \ldots,
        L^{(-\alpha-k)}_{m_k}(-z)\bigr) .
\end{split}
\end{equation}
\par
%
%
{}From the results of Subsecs.~IIIA and IIIB, we know that for the four cases mentioned above, namely $H_{a}$, $H_{b}$, $H_{c}$, and $H_{d}$, $H_x$ possesses ladder operators associated with a polynomial Heisenberg algebra. On the other hand, in Sec.~II, we presented the well-known ladder operators and polynomial Heisenberg algebras for the harmonic and radial oscillators. From the general construction of Subsec.~IVA, we obtain the integrals of motion as products of these ladder operators. We take $a_x$ as given by Eq.~(\ref{eq:ladder-k-HO-1}) for Hamiltonians $H_{a}$ and $H_{c}$, and as given by Eq.~(\ref{eq:ladder-k-RHO-1}) for $H_{b}$ and $H_{d}$. Moreover, $a_y$ is the usual ladder operator of the harmonic oscillator given by Eq.~(\ref{eq:ladder-HO}) for $H_{a}$ and $H_{b}$, and the usual ladder operator of the radial oscillator given by Eq.~(\ref{eq:ladder-RHO}) (with $k=0$) for $H_{c}$ and $H_{d}$. In all cases, $\lambda_x=2m_k +2$ and $\lambda_y=2$ and correspondingly $n_1=1$ and $n_2= m_k +1$, so that the order of the resulting integrals $I_{+}$ and $I_{-}$ is, in the four cases, $2m_{k}+2$, $3m_{k}+3$, $3m_{k}+3$, and $4m_{k}+4$, respectively. \par
%
%
The polynomial algebra can be determined explicitly using Eq.~(\ref{eq:algebra}). The polynomial $Q$ is given by Eq.~(\ref{eq:Q}) for $H_{a}$ and $H_{c}$ and by Eq.~(\ref{eq:Q-bis}) for $H_{b}$ and $H_{d}$. On the other hand, for the polynomial $S$ we use Eq.~(\ref{eq:Q-HO}) for $H_{a}$ and $H_{b}$ and Eq.~(\ref{eq:Q-RHO}) (with $k=0$) for $H_{c}$ and $H_{d}$. \par
%
%
The form of the potentials, the ladder operators and consequently the integrals of motion differs in the four cases. However, the structure of the energy spectrum is similar and also the action of the ladder operators $a_{x}$ and $a_{y}$ on the physical states $|\nu_{x}  \rangle_1$ and $|\nu_{y}  \rangle_2$, respectively. From unirreps of the polynomial algebras we can therefore perform an algebraic derivation of the energy spectrum and of the total degeneracies in an uniform manner for the four cases. \par
%
%
On using Eqs.~(\ref{eq:energy-add}) and (\ref{eq:partner-energy-bis}), we can calculate the energy spectrum directly. This yields  
\begin{equation}
  E_{i,N}=2N + \gamma_{i}, \qquad i=a,b,c,d, \label{eq:energ4cases}
\end{equation}
where $\gamma_{a}=0$, $\gamma_{b}=\alpha+k$, $\gamma_{c}=\alpha$, $\gamma_{d}= 2 \alpha + k$, and
\begin{equation}
  N=\nu_{x}+\nu_{y}+1 ,\qquad \nu_{x}=-m_{k}-1,\ldots,-m_{1}-1,0,1,2,\ldots, \qquad \nu_{y}=0,1,2\ldots. 
\end{equation}
The additive constant $\gamma_{i}$ in the energy eigenvalues does not affect the structure 
of the levels nor the degeneracies. In all cases a direct calculation leads to the  following results: 
\begin{equation}
  \deg(E_{N}) = \begin{cases}
      k-j+1  & \text{if $N=-m_{j},-m_j+1,\ldots,-m_{j-1}-1$ for $j=2,3,\ldots,k$}, \\ 
      k  & \text{if $N=-m_{1},-m_1+1,\ldots,0$}, \\ 
      N+k  & \text{if $N=1,2,3,\ldots$}. \\ 
  \end{cases} \label{eq:degen4cases}
 \end{equation}
Let us remark how the degeneracies differ from the well-known isotropic harmonic oscillator ones (i.e., for two harmonic oscillators in both axes), which are simply given by $N+1$. For the four families of systems considered here, the pattern is more complicated and there are bands of levels with degeneracies 1 to $k$ before the degeneracies increase as $N+k$. 
The width of these bands is related to parameters $m_{j}$ (with $j=1,\ldots,k$). We will show how even such a more complex structure can be deduced using the polynomial algebras and their finite-dimensional unirreps. \par
%
%
The eigenfunctions of the Hamiltonians $H_{a}$, $H_{b}$, $H_{c}$, and $H_{d}$ are given in terms of Hermite polynomials, Laguerre ones, type III Hermite EOP, or type III Laguerre EOP. In spite of this, they can be expressed in a uniform way. We can observe that on using $N=\nu_{x}+\nu_{y}+1$, the states $|\nu_{x} \rangle_{1} | \nu_{y} \rangle_{2}$ of the four 
systems can be written as
\begin{equation}
 |N, \nu_{x}  \rangle = |\nu_{x}  \rangle_{1} |N-\nu_{x}-1  \rangle_{2},  \label{eq:basis4cases}
\end{equation}
where the values of $N$ and $\nu_{x}$ are presented in Table I. We also list in this table the corresponding values of $\nu_{y}=N-\nu_{x}-1$ as they are used to calculate the action.\par
%
%
\begin{table}[h!]

\caption{States of superintegrable systems $H_{a}$, $H_{b}$, $H_{c}$, and $H_{d}$ in the $|N, \nu_{x}  \rangle$ basis with corresponding values of $\nu_{y}=N-\nu_{x}-1$. In this table, $j$ runs over 2, 3, \ldots, $k$.}

\begin{center}
\begin{tabular}{lll}
  \hline\hline\\[-0.2cm]
  $N$ & $\nu_{x}$ & $N-\nu_{x}-1$ \\[0.2cm]
  \hline\\[-0.2cm]
  $ -m_{j},-m_j+1,\ldots,$ & $-m_{k}-1,-m_{k-1}-1,\ldots,$ & $N+m_{k},N+m_{k-1},\ldots,$ \\[0.2cm]
      \quad $-m_{j-1}-1$ & \quad $-m_{j}-1$ & \quad $N+m_{j}$ \\[0.2cm]
  $ -m_{1},-m_1+1,\ldots,$ & $-m_{k}-1,-m_{k-1}-1,\ldots,$ & $ N+m_{k},N+m_{k-1},\ldots,$ \\[0.2cm]
      \quad 0 & \quad $-m_{1}-1 $ & \quad $ N+m_{1}$ \\[0.2cm]
  $1,2,\ldots$ & $-m_{k}-1,-m_{k-1}-1,\ldots,$ & $N+m_{k},N+m_{k-1},\ldots,$ \\[0.2cm]
  & \quad $-m_{1}-1,0,1,\ldots, N-1$ & \quad $N+m_{1},N-1,N-2,\ldots, 0$ \\[0.2cm] 
  \hline \hline
\end{tabular}
\end{center}

\end{table}
\par
%
%
The action of the integral $K$ associated with separation of variables is given by
\begin{equation}
 K |N, \nu_{x} \rangle = \frac{1}{2(m_{k}+1)}(2 \nu_{x}+1-N) |N, \nu_{x} \rangle . \label{eq:actionK4cases}
\end{equation}
That of the integrals $I_{+}$ and $I_{-}$ is calculated from the action of the ladder operators. We use iteratively  
Eqs.~(\ref{eq:action-HO}), (\ref{eq:action-RHO}), (\ref{eq:action-0})--(\ref{eq:action-2}), and~(\ref{eq:action-0-bis})--(\ref{eq:action-2-bis}) according to the values of $n_{1}$ and $n_{2}$. The states annihilated by $I_{+}$ are identified directly from $a_{y}^{n_{2}}|\nu_{y}  \rangle_{2}$ and those by  $I_{-}$ from $a_{x}^{n_{1}}|\nu_{x}  \rangle_{1}$.
The states annihilated by $I_{+}$ and $I_{-}$ are presented in  Tables VI and VII of Appendix B. They are given in terms of the $|N, \nu_{x}  \rangle$ basis. \par
%
%
{}For all other states, the action of the integrals $I_{+}$ and $I_{-}$ is given by
\begin{equation}
\begin{split}
 & I_{+}  |N,\nu_{x} \rangle = \alpha(N,\nu_{x},m_{k}) |N,\nu_{x}+ m_{k}+1 \rangle, \\
 & I_{-}  |N,\nu_{x} \rangle = \beta(N,\nu_{x},m_{k}) |N,\nu_{x}- m_{k}-1 \rangle. \label{eq:actionI4cases}
\end{split}
\end{equation}
The coefficients $\alpha(N,\nu_{x},m_{k})$ and $\beta(N,\nu_{x},m_{k})$ can be determined from the explicit action of the ladder operators. They differ in the four cases considered in this paper, but these explicit expressions do not play any role in the calculation 
of degeneracies using the finite-dimensional unirreps and we thus omit them. \par
%
%
{}Furthermore, as mentioned in Subsec. IVB, we also write all the states using the notation
$N=\lambda(m_{k}+1)+\mu$, $\mu=0,1,2,\ldots,m_{k}$. The negative values of $N$ can be obtained by taking the following values for $\lambda$ and $\mu$,
\begin{equation}
\begin{split}
  & N=-m_{j},-m_{j}+1,\ldots,-m_{j-1}-1, j = 2, 3, \ldots, k:  \\
  & \qquad  \lambda = -1, \quad \mu=m_{k}-m_{j}+1,m_{k}-m_{j}+2, \ldots, m_{k}-m_{j-1},\\
  & N=-m_{1},-m_{1}+1,\ldots,-1: \\
  & \qquad \lambda = -1, \quad \mu=m_{k}-m_{1}+1,m_k-m_1+2,\ldots,m_{k}.
\end{split}
\end{equation}
\par
%
%
We consider the states formed by the repeated action of the integral $I_{+}$ on the states $\xi$ annihilated by $I_{-}$, presented in Table VII. From the smallest integer $n+1$ such that
\begin{equation}
  (I_{+})^{n+1} \xi=0, 
\end{equation}
we deduce that $(I_{+})^n\xi$ is one of the zero modes of $I_{+}$ of Table VI, denoted by $\chi$. This process allows us to obtain the integer or half-integer $s$ (given by $\frac{n}{2}$) such that Eq.~(\ref{eq:rep}) is satisfied. A value of $\sigma$ is assigned to each member of this finite sequence. In this way, we calculate the values of $s$ realized for every $N$. In Table II, we list the values of $\lambda$, $\mu$, and $s$, together with the associated number $\cal N$ of unirreps per level, and the total level degeneracy.  \par
%
%
The results agree with those previously obtained for $k=1$ using the Daskaloyannis approach and presented in Table I of Ref.~\cite{marquette13c}. For this special value of $k$, lines 1, 2, 5, 6, 9, and 10 of Table II would be missing in the present approach. In order to show the results coincide for both methods, we also need to take into account the following relations between the parameters and integrals:
\begin{equation}
s=\frac{p}{2}, \qquad I_{0}=N-\frac{p}{2}, \qquad c=-u-\frac{p}{2}. \label{eq:rel4cases}
\end{equation}
\par
%
%
Note in addition that for the next value of $k$, namely $k=2$, lines 2, 6, and 10 would be missing in Table II. \par
%
%
\begin{table}[h!]

\caption{Set of $s$ values with their number of occurrences, number $\cal N$ of unirreps per level, and total level degeneracy for the polynomial algebra (\ref{eq:algebra}) corresponding to Hamiltonian $H_a$, $H_b$, $H_c$, or $H_d$. In this table, $j$ runs over 2, 3, \ldots, $k-1$.}

\begin{center}
\begin{tabular}{lllll}
  \hline\hline\\[-0.2cm]
  $\lambda$ & $\mu$ & $s$ & $\cal N$ & $\deg(E_N)$\\[0.2cm]
  \hline\\[-0.2cm]
  $-1$ & $1, \ldots, m_{k}-m_{k-1}$ & 0 & 1 & 1 \\[0.2cm]
  $-1$ & $m_{k}-m_{j}+1,\ldots,m_{k}-m_{j-1}$ & $0^{k-j+1}$ & $k-j+1$ & $k-j+1$ \\[0.2cm]
  $-1$ & $m_{k}-m_{1}+1,\ldots, m_{k}$ & $0^{k}$ & $k$ & $k$  \\[0.2cm]
  0 & 0 & $0^{k}$ & $k$ & $k$  \\[0.2cm]
  $0$ & $1,\ldots,m_{k}-m_{k-1}$ & $\frac{1}{2}$ & $\mu+k-1$ & $N+k$ \\[0.2cm]
  & & $0^{\mu+k-2}$ & & \\[0.2cm]
  $0$& $m_{k}-m_{j}+1, \ldots,m_{k}-m_{j-1}$ & $(\frac{1}{2})^{k-j+1}$ & $\mu+j-1$ & $N+k$ \\[0.2cm]
     &  & $0^{\mu-k+2j-2}$ & & \\[0.2cm]
  $0$& $m_{k}-m_{1}+1, \ldots,m_{k}$ & $(\frac{1}{2})^{k}$ & $\mu$ & $N+k$ \\[0.2cm]
     &  & $0^{\mu-k}$ & & \\[0.2cm] 
  $1,2,\ldots$ & 0 & $(\frac{\lambda}{2})^k$ & $m_k+1$ & $N+k$ \\[0.2cm]
     & & $(\frac{\lambda-1}{2})^{m_k-k+1}$ & & \\[0.2cm]
  $1,2,\ldots$& $1,\ldots,m_{k}-m_{k-1}$ & $\frac{\lambda+1}{2}$ & $m_{k}+1$ & $N+k$ \\[0.2cm]
     &  & $(\frac{\lambda}{2})^{\mu+k-2}$ & & \\[0.2cm]
     &  & $(\frac{\lambda-1}{2})^{m_{k}-\mu-k+2}$ & & \\[0.2cm]
  $1,2,\ldots$& $m_{k}-m_{j}+1,\ldots,m_{k}-m_{j-1}$ & $(\frac{\lambda+1}{2})^{k-j+1}$ & $m_{k}+1$ & $N+k$ \\[0.2cm]
     &  & $(\frac{\lambda}{2})^{\mu-k+2j-2}$ & & \\[0.2cm]
     &  & $(\frac{\lambda-1}{2})^{m_{k}-\mu-j+2}$ & & \\[0.2cm]
  $1,2,\ldots$& $m_{k}-m_{1}+1,\ldots,m_{k}$ & $(\frac{\lambda+1}{2})^k$ & $m_{k}+1$ & $N+k$ \\[0.2cm]
     &  & $(\frac{\lambda}{2})^{\mu-k}$ & & \\[0.2cm]
     &  & $(\frac{\lambda-1}{2})^{m_{k}-\mu+1}$ & & \\[0.2cm]    
  \hline \hline
\end{tabular}
\end{center}

\end{table}
\par
%
%
\subsection{\boldmath Combination of two $k$-step rationally-extended potentials}

Let us consider 2D Hamiltonians of the type (\ref{eq:H}) with a $k$-step extension of the harmonic oscillator, given by Eq.~(\ref{eq:partner-add}), and/or a $k$-step extension of the radial oscillator, given by Eq.~(\ref{eq:partner-add-bis}), in both axes. We can form three new infinite families of superintegrable systems, which include some of those introduced 
in previous papers \cite{marquette13b,marquette13c}:
\begin{equation}
\begin{split}
  H_{e} &= - \frac{d^2}{dx^2} - \frac{d^2}{dy^2} +  x^2 + y^2 - 2k_{1} - 2k_2 \\
  & \quad - 2 \frac{d^2}{dx^2} \log {\cal W}({\cal H}_{m_1}, {\cal H}_{m_2}, \ldots, {\cal H}_{m_{k_{1}}})  
      - 2 \frac{d^2}{dy^2} \log {\cal W}({\cal H}_{n_1}, {\cal H}_{n_2}, \ldots,{\cal H}_{n_{k_{2}}}),
\end{split}
\end{equation}
\begin{equation}
\begin{split}
  H_{f} &=  - \frac{d^2}{dx^2} - \frac{d^2}{dy^2} +  \frac{1}{4} x^2 + \frac{l(l+1)}{x^2} +  \frac{1}{4} y^2 + \frac{l(l+1)}{y^2} 
      - k_{1} -k_2 \\
  & \quad - 2 \frac{d^2}{dx^2} \log \tilde{\cal W}\bigl(L^{(-\alpha-k_{1})}_{m_1}(-z), L^{(-\alpha-k_{1})}_{m_2}(-z), 
      \ldots,L^{(-\alpha-k_{1})}_{m_{k_{1}}}(-z)\bigr)  \\ 
  & \quad - 2 \frac{d^2}{dy^2} \log \tilde{\cal W}\bigl(L^{(-\alpha-k_{2})}_{n_1}(-z), L^{(-\alpha-k)}_{n_2}(-z), \ldots, 
      L^{(-\alpha-k_{2})}_{n_{k_{2}}}(-z)\bigr),
\end{split}
\end{equation}
\begin{equation}
\begin{split}
  H_{g} &=  - \frac{d^2}{dx^2} - \frac{d^2}{dy^2} +  x^2 +  \frac{1}{4} y^2 + \frac{l(l+1)}{y^2} - 2k_{1} -k_2 \\
  & \quad - 2 \frac{d^2}{dx^2} \log {\cal W}({\cal H}_{m_1}, {\cal H}_{m_2}, \ldots, {\cal H}_{m_{k_{1}}})   \\ 
  & \quad- 2 \frac{d^2}{dy^2} \log \tilde{\cal W}\bigl(L^{(-\alpha-k_{2})}_{n_1}(-z), L^{(-\alpha-k_{2})}_{n_2}(-z), \ldots, 
       L^{(-\alpha-k_{2})}_{n_{k_{2}}}(-z)\bigr).
\end{split}
\end{equation}
\par
%
%
By construction, these systems are superintegrable for any $k_{1}$ and $k_{2}$. We take $a_x$ as given by Eq.~(\ref{eq:ladder-k-HO-1}) for $H_{e}$ and $H_{g}$, and as given by Eq.~(\ref{eq:ladder-k-RHO-1}) for $H_{f}$, while for $a_y$ we assume Eq.~(\ref{eq:ladder-k-HO-1}) for $H_{e}$ and Eq.~(\ref{eq:ladder-k-RHO-1}) for $H_{f}$ and $H_{g}$. In all cases, we have $\lambda_x=2m_{k_{1}} +2$ and $\lambda_y=2 n_{k_{2}} +2 $, so that $n_1=n_{k_{2}} +1$ and $n_2= m_{k_{1}} +1$. The order of the integrals $I_{+}$ and $I_{-}$ is $2(m_{k_{1}}+1)(n_{k_{2}}+1)$,  $4(m_{k_{1}}+1)(n_{k_{2}}+1)$, and  $3(m_{k_{1}}+1)(n_{k_{2}}+1)$ for these three cases, respectively. The polynomial algebra can be determined explicitly using Eq.~(\ref{eq:algebra}). In this equation, the polynomial $Q$ is given by  Eq.~(\ref{eq:Q}) for $H_{e}$ and $H_{g}$  and by Eq.~(\ref{eq:Q-bis}) for $H_{f}$. On the other hand, for the polynomial $S$ we use Eq.~(\ref{eq:Q}) for $H_{e}$ and Eq.~(\ref{eq:Q-bis}) for $H_{f}$ and $H_{g}$ . \par
%
%
{}For the purpose of an algebraic derivation of the degeneracies, let us 
consider the particular case of two one-step extensions with $m_1$ and $n_1$ (without lost of generality we may assume $m_1 \geq n_1$).
Only the case when $m_1=n_1=m$ has been published \cite{marquette13c} and there are also some unpublished results for $m_1=4$, $m_2=2$, but the general problem is difficult to solve in the Daskaloyannis approach. Equations (\ref{eq:energy-add}) and (\ref{eq:partner-energy-bis}) lead to the following degenerate energy spectrum:     
\begin{equation}
  E_{i,N}=2N + \delta_{i}, \qquad i=e,f,g,  \label{eq:energy3cases}
\end{equation}
where $\delta_{e}=0$, $\delta_{f}=2\alpha + 2$, $\delta_{g}=\alpha +1$, and 
\begin{equation}
  N=\nu_{x}+\nu_{y}+1, \qquad \nu_x = -m_{1}-1, 0, 1, 2, \ldots, \qquad \nu_y = -n_{1}-1, 0, 1, 2, \ldots. 
\end{equation}
The degeneracies can be written in the three cases as
\begin{equation}
  \deg(E_N) = \begin{cases}
    1 & \text{if $N = -m_{1}-n_{1}-1, -m_1, -m_1+1, \ldots,-n_{1}-1$},\\
    2 & \text{if $N = -n_{1}, -n_1+1, \ldots,0$},\\
    N+2 & \text{if $N = 1, 2, \ldots$.} 
  \end{cases}  \label{eq:degen3cases} 
\end{equation}
In these cases, we note the presence of bands with the same total degeneracies as in the systems of Subsec.~IVC. The ladder operators of the $k$-step extensions of the harmonic and radial oscillators
act on the physical states $| \nu_{x} \rangle_{1} | \nu_{y} \rangle_{2}$ in a similar way and consequently we can use a uniform notation. Since $\nu_{y}=-n_{1}-1$ or $\nu_y \geq 0$, the relation $N=\nu_{x}+\nu_{y}+1$ implies that $\nu_{x}=N+n_{1}$ or $\nu_{x} \leq N-1$. The states can be rewritten as
\begin{equation}
  | N, \nu_{x} \rangle =   | \nu_{x} \rangle_{1} | N-\nu_{x} -1 \rangle_{2} ,
\end{equation}
with the corresponding values of $N$ and $\nu_{x}$ presented in Table III. We also list there the corresponding values of $\nu_{y}=N-\nu_{x}-1$ as they are useful to compute the action
of the integrals of motion and to identify the states they annihilate. \par
%
%
\begin{table}[h!]

\caption{States of superintegrable systems $H_e$, $H_f$, and $H_g$ (with $k_1=k_2=1$ and $m_1 \ge n_1$) in the $ | N, \nu_{x} \rangle$ basis and corresponding values of $\nu_{y}=N-\nu_{x}-1$.}

\begin{center}
\begin{tabular}{lll}
  \hline\hline\\[-0.2cm]
  $N$ & $\nu_{x}$ & $N-\nu_{x}-1$ \\[0.2cm]
  \hline\\[-0.2cm]
  $-m_{1}-n_{1}-1$ & $-m_{1}-1$ & $-n_{1}-1$ \\[0.2cm]
  $ -m_{1},-m_1+1,\ldots,-n_{1}-1$ & $-m_{1}-1$ & $ 0,1,\ldots, m_{1}-n_{1}-1$ \\[0.2cm]
  $ -n_{1},-n_1+1,\ldots,0$ & $-m_{1}-1, N+n_{1}$ & $N+m_{1},-n_{1}-1$ \\[0.2cm]
  $1,2,\ldots$ & $-m_{1}-1,0,1,\ldots,N-1,$ & $N+m_{1},N-1,N-2,\ldots,0,$ \\[0.2cm]
  & \quad $N+n_{1}$ & \quad $-n_{1}-1$ \\[0.2cm]  
  \hline \hline
\end{tabular}
\end{center}

\end{table}
\par
%
%
The action of the integral $K$ is given by the following equation
\begin{equation}
  K | N, \nu_{x} \rangle = \frac{1}{2(m_{1}+1)(n_{1}+1)}(2 \nu_{x} +1 -N) | N, \nu_{x} \rangle .
\end{equation}
That of the integrals $I_{+}$ and $I_{-}$ can be obtained directly from the action of the ladder operators given by Eqs.~(\ref{eq:action-0})--(\ref{eq:action-2}) and (\ref{eq:action-0-bis})--(\ref{eq:action-2-bis}). It is important to distinguish the states annihilated by $I_{+}$ and $I_{-}$, which are presented in Tables VIII and IX of Appendix B. The action of the integrals on all other states is given by 
\begin{equation}
\begin{split}
  & I_{+}  |N,\nu_{x} \rangle =\tilde{\alpha}(N,\nu_{x},m_{1},n_{1}) |N,\nu_{x}+ (m_{1}+1)(n_{1}+1) \rangle , \\
  & I_{-}  |N,\nu_{x} \rangle =\tilde{\beta}(N,\nu_{x},m_{1},n_{1})  |N,\nu_{x}- (m_{1}+1)(n_{1}+1) \rangle .  
\end{split} \label{eq:actionI3cases}
\end{equation}
The coefficients $\tilde{\alpha}(N,\nu_{x},m_{1},n_{1})$ and $\tilde{\beta}(N,\nu_{x},m_{1},n_{1})$ can be determined from the explicit action of the ladder operators. As they do not play any role in the derivation of the total degeneracies and would differ in the three cases, we omit them. The unirreps are calculated directly by acting repeatedly with $I_+$ on states annihilated by $I_{-}$ until we reach a state that is annihilated. Here we set $N=\lambda M+\mu$, $M=(m_{1}+1)(n_{1}+1)$, $\mu=0,1,\ldots,m_{1}n_{1}+m_{1}+n_{1}$. The finite-dimensional unirreps are listed in Table IV. \par
%
%
\begin{table}[h!]

\caption{Set of $s$ values with their number of occurrences, number $\cal N$ of unirreps per level, and total level degeneracy for the polynomial algebra (\ref{eq:algebra}) corresponding to Hamiltonian $H_e$, $H_f$, or $H_g$ with $k_1=k_2=1$ and $m_1 \ge n_1$.}

\begin{center}
\begin{tabular}{lllll}
  \hline\hline\\[-0.2cm]
  $\lambda$ & $\mu$ & $s$ & $\cal N$ & $\deg(E_N)$\\[0.2cm]
  \hline\\[-0.2cm]
  $-1$ & $m_{1}n_{1}$ & 0 & 1 & 1 \\[0.2cm]
  $-1$ & $m_{1}n_{1}+n_{1}+1,\ldots,m_{1}n_{1}+m_{1}$ & $0$ & $1$ & $1$ \\[0.2cm]
  $-1$ & $m_{1}n_{1}+m_{1}+1,\ldots,m_{1}n_{1}+m_{1}+n_{1}$ & $0^{2}$ & $2$ & $2$ \\[0.2cm]
  0 & $0,\ldots, m_{1}n_{1}-1$ & $0^{\mu+2}$ & $\mu+2$ & $N+2$  \\[0.2cm]
  $0$ & $m_{1}n_{1}$ & $\frac{1}{2}$ & $\mu+1$ & $N+2$ \\[0.2cm]
      & & $0^{\mu}$ & & \\[0.2cm]
  $0$& $m_{1}n_{1}+1, \ldots,m_{1}n_{1}+n_{1}$ & $0^{\mu+2}$ & $\mu+2$ & $N+2$ \\[0.2cm]
  $0$& $m_{1}n_{1}+n_{1}+1, \ldots,m_{1}n_{1}+m_{1}$ & $\frac{1}{2}$ & $\mu+1$ & $N+2$ \\[0.2cm]
      &  & $0^{\mu}$ & & \\[0.2cm] 
  $0$& $m_{1}n_{1}+m_{1}+1, \ldots,m_{1}n_{1}+m_{1}+n_{1}$ & $(\frac{1}{2})^2$ & $\mu$ & $N+2$ \\[0.2cm]
      &  & $0^{\mu-2}$ & & \\[0.2cm] 
  $1,2,\ldots$& $0,\ldots,m_{1}n_{1}-1$ & $(\frac{\lambda}{2})^{\mu+2}$ & $M$ & $N+2$ \\[0.2cm]
      &  & $(\frac{\lambda-1}{2})^{M-\mu-2}$ & & \\[0.2cm]
  $1,2,\ldots$& $m_{1}n_{1}$ & $\frac{\lambda+1}{2}$ & $M$ & $N+2$ \\[0.2cm]
      &  & $(\frac{\lambda}{2})^{\mu}$ & & \\[0.2cm]
      &  & $(\frac{\lambda-1}{2})^{M-\mu-1}$ & & \\[0.2cm]
  $1,2,\ldots$& $m_{1}n_{1}+1,\ldots,m_{1}n_{1}+n_{1}$ & $(\frac{\lambda}{2})^{\mu+2}$ & $M$ & $N+2$ \\[0.2cm]
      &  & $(\frac{\lambda-1}{2})^{M-\mu-2}$ & & \\[0.2cm]
  $1,2,\ldots$& $m_{1}m_{2}+n_{1}+1,\ldots,m_{1}n_{1}+m_{1}$ & $\frac{\lambda+1}{2}$ & $M$ & $N+2$ \\[0.2cm]
      &  & $(\frac{\lambda}{2})^{\mu}$ & & \\[0.2cm]
      &  & $(\frac{\lambda-1}{2})^{M-\mu-1}$ & & \\[0.2cm]
  $1,2,\ldots$& $m_{1}n_{1}+m_{1}+1,\ldots,m_{1}n_{1}+m_{1}+n_{1}$ & $(\frac{\lambda+1}{2})^{2}$ & $M$ & $N+2$ 
           \\[0.2cm]
      &  & $(\frac{\lambda}{2})^{\mu-2}$ & & \\[0.2cm]
      &  & $(\frac{\lambda-1}{2})^{M-\mu}$ & & \\[0.2cm]  
  \hline \hline
\end{tabular}
\end{center}

\end{table}
\par
%
%
The results agree with those previously obtained for $m_{1}=n_{1}=m$ and for $m_{1}=4$, $n_{1}=2$ by using Daskaloyannis method of calculating the finite-dimensional unirreps of superintegrable systems. In the case $m_{1}=n_{1}=m$, for instance, we have $N=\lambda (m+1)^{2}+\mu$,  $\mu=0,1,\ldots,m(m+2)$ in the present approach. In Table II of Ref.~\cite{marquette13c}, on the other hand, we used the notation $\mu = \rho(m+1) + \sigma$ with $\rho$ and $\sigma$ running over 0, 1, \ldots, $m$ (note that this $\sigma$ has nothing to do with $\sigma$ introduced in Eq.~(\ref{eq:basisunirreps})). The correspondence between the $\mu$ and $(\rho, \sigma)$ values is given in Table V. On taking it into account, it can be easily checked that the results of Ref.~\cite{marquette13c} are recovered as a special case of Table IV. \par
%
%
\begin{table}[h!]

\caption{Comparaison with $N=\lambda(m+1)^{2}+\mu$ where $\mu=\rho(m+1)+\sigma$}

\begin{center}
\begin{tabular}{lll}
  \hline\hline\\[-0.2cm]
     $\mu$ & $\rho$ & $\sigma$ \\[0.2cm]
  \hline\\[-0.2cm]
     $0, 1, \ldots,  m^{2}-1$  & $0,1,\ldots,m-2$ &  $0,1,\ldots,m$ \\[0.2cm]
     $0, 1, \ldots,  m^{2}-1$   & $m-1$  & $0$ \\[0.2cm]
     $m^{2}$   & $m-1$  & 1 \\[0.2cm]
     $ m^{2}+1, m^2+2,\ldots,m^{2}+m$   & $m-1$  & $2,3,\ldots,m$ \\[0.2cm]
     $m^{2}+1, m^2+2,\ldots,m^{2}+m$   & $m$  & 0 \\[0.2cm]
     $m^{2}+m+1,m^2+m+2,\ldots,m^{2}+2m$   & $m$  & $1,2,\ldots,m$ \\[0.2cm]     
  \hline \hline
\end{tabular}
\end{center}

\end{table}
\par
\clearpage
%
%
\section{CONCLUSION}

In this paper, for $k$-step extensions of the harmonic and radial oscillators related with type III EOP, we obtained new ladder operators that satisfy 
polynomial Heisenberg algebras with only infinite-dimensional unirreps. The construction is based on supersymmetric quantum mechanics and relies on a combination of the state-adding and state-deleting approaches. These results extend those obtained recently for one-step extensions of the harmonic oscillator \cite{marquette13c}. 
\par
%
%
Moreover, from these ladder operators we generated seven new infinite families of superintegrable systems with higher-order integrals of motion. 
These families generalize various superintegrable systems discovered by Gravel \cite{gravel04}. It is interesting to note that one- and two-step extensions  of the harmonic oscillator (for specific values of $m_{1}$ and $m_{2}$) are also related with quantum systems involving the fourth Painlev\'e transcendent \cite{gravel04,marquette09b,marquette10}.
The new ladder operators are not only significant in order to construct integrals of motion, but more importantly allow to obtain a higher-order polynomial algebra, from which an algebraic derivation of the whole energy spectrum and the 
total degeneracies can be done by using finite-dimensional unirreps. Such ladder operators are not of the lowest possible order in contrast with standard ladder operators obtained 
from supersymmetric quantum mechanics. The latter, however, admit $k$ singlet states and, as a consequence, do not allow to create integrals of motion with a corresponding polynomial
algebra whose finite-dimensional unirreps give rise to the whole energy spectrum and the total degeneracies. 
\par
%
%
Another novelty of the paper is the introduction of a different method than the Daskaloyannis approach used in previous works for the calculation of finite-dimensional unirreps. It takes advantage of the fact we know the underlying structure of the integrals. 
Tables II and IV list the obtained unirreps and degeneracies. We recover as particular cases results obtained from the Daskaloyannis approach \cite{marquette13c}. A very interesting aspect of our results is the fact that by allowing $k$-step extensions of the harmonic or radial oscillator we can create more complex patterns for the degeneracies. The seven families discussed here have bands of levels with the same amount of total degeneracies. The length of these bands depends on the choice of the Hamiltonian parameters $m_{i}$ and $n_{j}$. This may open 
the way of another type of spectral design, i.e., the possibility of creating systems with a given type of degeneracies. These systems and their new ladder operators themselves could have applications in other contexts. Let us mention that the anisotropic oscillator and its related degeneracies
found applications in nuclear physics \cite{bonatsos}.  
\par
%
%
Let us make further remarks and describe how these results could also be generalized in various ways. One possible avenue is the extension to 3D systems. It would be of great interest to study their degeneracies through finite-dimensional unirreps. We also restricted our algebraic treatment to isotropic cases, but generalization to anisotropic ones could be investigated. Furthermore, from Ref.~\cite{marquette13a}, we know that two-step extensions of the harmonic oscillator allow not only ladder operators with two singlet states, but also ladder operators with a doublet state. We showed in this paper there exist ladder operators with only infinite sequences of levels for $k$-step extensions of the harmonic oscillator that differ from the standard ones with $k$ singlets. The study of other types of ladder operators and, in particular, those with multiplet states for $k$-step extensions needs to be done.
\par
%
%
\section*{ACKNOWLEDGMENTS}

C.\ Q.\ would like to thank G.\ Pogosyan for an interesting discussion. The research of I.\ M.\ was supported by the Australian Research Council through Discovery Project No.\ DP110101414 and Discovery Early Career Researcher Award DE130101067.
I.\ M.\ thanks the Universit\'e Libre de Bruxelles and the CRM for their hospitality. \par
%
%
\section*{\boldmath APPENDIX A: GOING FROM $H^{(1)}$ TO $\bar{H}^{(1)}$ IN THE RADIAL HARMONIC OSCILLATOR CASE}

\renewcommand{\theequation}{A.\arabic{equation}}
\setcounter{equation}{0}

In addition to the two sets  of partner Hamiltonians $(H^{(1)}, H^{(2)})$ and $(\bar{H}^{(1)}, \bar{H}^{(2)})$ considered in Subsec.~IIB, we plan to introduce here a third one $(\tilde{H}^{(1)}, \tilde{H}^{(2)})$, such that $\tilde{H}^{(1)} = H^{(1)}$ while $\tilde{H}^{(2)}$ only differs from $\bar{H}^{(1)}$ by some additive constant.\par
%
%
This can be achieved by considering $n$th-order differential operators $\tilde{\cal A}$ and $\tilde{\cal A}^{\dagger}$ intertwining with $\tilde{H}^{(1)}$ and $\tilde{H}^{(2)}$ as in Eq.~(\ref{eq:intertwine}) and constructed from $n=m_k+1$ polynomial-type seed solutions of $\tilde{H}^{(1)} = H^{(1)}$ of class II, i.e., $(\varphi_1, \varphi_2, \ldots, \varphi_n) \to \bigl(\tilde{\phi}^{l+k}_0, \tilde{\phi}^{l+k}_1, \ldots, \tilde{\phi}^{l+k}_{m_k}\bigr)$, where \cite{cq11a}
\begin{equation}
\begin{split}
  & \tilde{\phi}^{l+k}_j(x) = \chi^{\rm II}_{l+k}(z) L^{(-\alpha-k)}_j(z), \qquad j=0, 1, \ldots, m_k, \\
  & \chi^{\rm II}_{l+k}(z) = z^{-\frac{1}{4}(2\alpha+2k-1)} e^{-\frac{1}{2}z} = z^{-\alpha-k} \eta_{l+k}(z).
\end{split}  \label{eq:seed-II}
\end{equation}
The corresponding eigenvalue of $\tilde{\phi}^{l+k}_j(x)$ is given by $E^{\rm II}_{l+k,j} = - \alpha - k + 2j + 1$.\par
%
%
The two partner potentials now read
\begin{equation}
\begin{split}
  \tilde{V}^{(1)}(x) &= V^{(1)}(x) = V_{l+k}(x), \\
  \tilde{V}^{(2)}(x) &= \tilde{V}^{(1)}(x) - 2 \frac{d^2}{dx^2} \log {\cal W}\bigl(\tilde{\phi}^{(l+k)}_0(x), \tilde{\phi}^{(l+k)}_1(x),  
       \ldots, \tilde{\phi}^{(l+k)}_{m_k}(x)\bigr), 
\end{split}
\end{equation}
where the latter is nonsingular provided $\alpha + k > m_k$. As from standard properties of Wronskians \cite{muir}, it can be shown that ${\cal W}\bigl(\tilde{\phi}^{(l+k)}_0, \tilde{\phi}^{(l+k)}_1, \ldots, \tilde{\phi}^{(l+k)}_{m_k}\bigr) \propto \bigl(\chi^{\rm II}_{l+k}(z)\bigr)^{m_k+1} z^{m_k(m_k+1)/4}$, it is straightforward to get the relation
\begin{equation}
  \tilde{V}^{(2)}(x) = V_{l+k-m_k-1}(x) + m_k + 1 = \bar{V}^{(1)}(x) + m_k + 1,  \label{eq:V-tilde-2-bis}
\end{equation}
which establishes the announced property.\par
%
%
As a check, we may compare the bound-state energies and wavefunctions of $\tilde{V}^{(2)}(x)$, obtained from those of $\tilde{V}^{(1)}(x)$ in $(m_k+1)$th-order SUSYQM, with those of $\bar{V}^{(1)}(x)$ given in Subsec.~IIB. Since type II seed functions lead to isospectral transformations, Eq.~(\ref{eq:RHO-E}) yields
\begin{equation}
  \tilde{E}^{(2)}_{l+k-m_k-1,\nu} = \tilde{E}^{(1)}_{l+k,\nu} = E^{(1)}_{l+k,\nu} = 2\nu + \alpha + k + 1, \qquad \nu=0, 1, 2, \ldots,
\end{equation}
which is compatible with Eqs.~(\ref{eq:E-bar}) and (\ref{eq:V-tilde-2-bis}). Furthermore, in the wavefunctions
\begin{equation}
  \tilde{\psi}^{(2)}_{l+k-m_k-1,\nu}(x) \propto \frac{{\cal W}\bigl(\tilde{\phi}^{(l+k)}_0, \tilde{\phi}^{(l+k)}_1, \ldots, 
  \tilde{\phi}^{(l+k)}_{m_k}, \psi^{(l+k)}_{\nu}\bigr)}{{\cal W}\bigl(\tilde{\phi}^{(l+k)}_0, \tilde{\phi}^{(l+k)}_1, \ldots, 
  \tilde{\phi}^{(l+k)}_{m_k}\bigr)}, \qquad \nu=0, 1, 2, \ldots, 
\end{equation}
the denominator is already known, while, on using Eq.~(\ref{eq:seed-II}), the numerator can be readily shown to be given by
\begin{equation}
\begin{split}
  & {\cal W}\bigl(\tilde{\phi}^{(l+k)}_0, \tilde{\phi}^{(l+k)}_1, \ldots, \tilde{\phi}^{(l+k)}_{m_k}, \psi^{(l+k)}_{\nu}\bigr) \\
  & \qquad \propto \bigl(\chi^{\rm II}_{l+k}\bigr)^{m_k+2} z^{\alpha+k + \frac{1}{4}(m_k-2)(m_k+1)} 
        L^{(\alpha+k-m_k-1)}_{\nu}(z).
\end{split}
\end{equation}
Hence
\begin{equation}
\begin{split}
  & \tilde{\psi}^{(2)}_{l+k-m_k-1,\nu}(x) \propto \chi^{\rm II}_{l+k}(z) z^{\alpha+k-\frac{1}{2}(m_k+1)} 
       L^{(\alpha+k-m_k-1)}_{\nu}(z) \\
  & \qquad \propto \eta_{l+k-m_k-1}(z) L^{(\alpha+k-m_k-1)}_{\nu}(z) \propto \bar{\psi}^{(1)}_{l+k-m_k-1,\nu}(x), 
\end{split}
\end{equation}
which completes the check.\par
%
%
\section*{\boldmath APPENDIX B: STATES ANNIHILATED BY THE INTEGRALS $I_{+}$ AND $I_{-}$ IN THE BASIS $|N, \nu_{x} \rangle$}

\renewcommand{\theequation}{B.\arabic{equation}}
\setcounter{equation}{0}

We present in Tables VI and VII the states annihilated by the integrals $I_{+}$ and $I_{-}$ in the case of the four Hamiltonians of Subsec.~IVC. They are given in the $|N, \nu_{x} \rangle$ basis.\par
%
%
\begin{table}[h!]

\caption{The states $\chi$ annihilated by the integrals $I_{+}$ in the $|N, \nu_{x} \rangle$ basis for the superintegrable systems $H_a$, $H_b$, $H_c$, and $H_d$.}

\begin{center}
\begin{tabular}{ll}
  \hline\hline\\[-0.2cm]
    $N$ & $\nu_{x}$  \\[0.2cm]
  \hline\\[-0.2cm]
    $-m_{j},-m_j+1,\ldots,-m_{j-1}-1$, & $-m_{k}-1,-m_{k-1}-1,\ldots,-m_{j}-1$  \\[0.2cm]
        \quad  $j=2,3,\ldots,k$ & \\[0.2cm]
    $-m_{1},-m_1+1,\ldots,0$ &  $-m_{k}-1,-m_{k-1}-1,\ldots,-m_{1}-1$  \\[0.2cm]
    $1,2,\ldots, m_{k}-m_{k-1}$ & $-m_{k-1}-1,-m_{k-2}-1, \ldots,-m_{1}-1,$ \\[0.2cm]
        & \quad $0,1,\ldots,N-1$ \\[0.2cm]
    $m_{k}-m_{j}+1,m_k-m_j+2,\ldots, m_{k}-m_{j-1},$ & $-m_{j-1}-1,-m_{j-2}-1, \ldots,-m_{1}-1,$  \\[0.2cm]
        \quad $j=2,3, \ldots, k-1$ & \quad $0,1,\ldots,N-1$ \\[0.2cm]
    $m_{k}-m_{1}+1,m_k-m_1+2,\ldots, m_{k}+1$  & \quad $0,1,\ldots,N-1$  \\[0.2cm] 
    $m_{k}+2,m_k+3,\ldots$  & $N-m_{k}-1,N-m_{k},\ldots,N-1$  \\[0.2cm] 
   
  \hline \hline
\end{tabular}
\end{center}

\end{table}
\par
%
%
\begin{table}[h!]

\caption{The states $\xi$ annihilated by the integrals $I_{-}$ in the $|N, \nu_{x} \rangle$ basis for the superintegrable systems $H_a$, $H_b$, $H_c$, and $H_d$.}

\begin{center}
\begin{tabular}{ll}
  \hline\hline\\[-0.2cm]
    $N$ & $\nu_{x}$  \\[0.2cm]
  \hline\\[-0.2cm]
    $-m_{j},-m_{j}+1,\ldots,-m_{j-1}-1$,  & $-m_{k}-1,-m_{k-1}-1,\ldots,-m_{j}-1$  \\[0.2cm]
         \quad $j=2,3,\ldots,k$ & \\[0.2cm]
    $-m_{1},-m_1+1,\ldots,0$ &  $-m_{k}-1,-m_{k-1}-1,\ldots,-m_{1}-1$  \\[0.2cm]
    $1,2,\ldots, m_{k}-m_{k-1}$ & $-m_{k}-1,-m_{k-1}-1,\ldots,-m_{1}-1,$ \\[0.2cm]
         & \quad $1,2,\ldots,N-1$ \\[0.2cm]
    $m_{k}-m_{j}+1,m_k-m_j+2,\ldots,$ & $-m_{k}-1,-m_{k-1}-1,\ldots,-m_{1}-1,1,2,\ldots,$ \\[0.2cm]
         \quad $m_{k}-m_{j-1}, j=2,3, \ldots,k-1$ & \quad $m_{k}-m_{k-1}-1,m_{k}-m_{k-1}+1,\ldots,$\\[0.2cm]
          & \quad $m_{k}-m_{j}-1,m_{k}-m_{j}+1,\ldots,N-1$  \\[0.2cm]
    $m_{k}-m_{1}+1,m_k-m_1+2,\ldots,m_{k}+1$  & $-m_{k}-1,-m_{k-1}-1,\ldots,-m_{1}-1,1,2,\ldots,$\\[0.2cm]
         & $\quad m_{k}-m_{k-1}-1,m_{k}-m_{k-1}+1,\ldots,$\\[0.2cm]
         & \quad $m_{k}-m_{1}-1,m_{k}-m_{1}+1,\ldots,N-1$  \\[0.2cm] 
    $m_{k}+2,m_k+3,\ldots$  & $-m_{k}-1,-m_{k-1}-1,\ldots,-m_{1}-1,1,2,\ldots,$  \\[0.2cm] 
         & $\quad m_{k}-m_{k-1}-1,m_{k}-m_{k-1}+1,\ldots,,$ \\[0.2cm]
         & $\quad m_{k}-m_{1}-1,m_{k}-m_{1}+1,\ldots,m_{k}$  \\[0.2cm]   
  \hline \hline
\end{tabular}
\end{center}

\end{table}
\par
%
%
We present in Tables VIII and IX the states annihilated by the integrals $I_{+}$ and $I_{-}$ in the case of the three Hamiltonians of Subsec.~IVD. They are given in the $|N, \nu_{x} \rangle$ basis.
\begin{table}[h!]

\caption{The states $\chi$ annihilated by the integrals $I_{+}$ in the $|N, \nu_{x} \rangle$ basis for the superintegrable systems $H_e$, $H_f$, and $H_g$ with $k_1=k_2=1$ and $m_1 \ge n_1$.}

\begin{center}
\begin{tabular}{ll}
  \hline\hline\\[-0.2cm]
    $N$ & $\nu_{x}$  \\[0.2cm]
  \hline\\[-0.2cm]
    $-m_{1}-n_{1}-1,-m_{1},-m_1+1,\ldots,$ & $-m_{1}-1$  \\[0.2cm]
        \quad $-n_{1}-1$ & \\[0.2cm]
    $-n_{1},-n_1+1,\ldots,0$ & $-m_{1}-1,N+n_{1}$  \\[0.2cm]
    $1,2,\ldots,m_{1}n_{1}-1$ & $-m_{1}-1,0,1,\ldots,N-1,N+n_{1}$  \\[0.2cm]
    $m_{1}n_{1}$ & $0,1,\ldots,m_{1}n_{1}-1,(m_{1}+1)n_{1}$  \\[0.2cm]
    $m_{1}n_{1}+1,m_1n_1+2,\ldots,$ & $-m_{1}-1,0,1,\ldots,N-1,N+n_{1}$  \\[0.2cm]
        \quad $(m_{1}+1)n_{1}$ & \\[0.2cm]
    $m_{1}n_{1}+n_{1}+1,m_{1}n_{1}+n_{1}+2,\ldots,$ & $0,1,\ldots,N-1,N+n_{1}$  \\[0.2cm]
        \quad $m_{1}(n_{1}+1)$ & \\[0.2cm]
    $m_{1}(n_{1}+1)+1,m_1(n_1+1)+2,\ldots,$ & $0,1,\ldots,N-m_{1}(n_{1}+1)-2,N-m_{1}(n_{1}+1),$  \\[0.2cm]
        \quad $(m_{1}+1)(n_{1}+1)$ & \quad $N-m_{1}(n_{1}+1)+1,\ldots,N-1,N+n_{1}$  \\[0.2cm]
     $(m_{1}+1)(n_{1}+1)+1,$ & $N-(m_{1}+1)(n_{1}+1),N-(m_{1}+1)(n_{1}+1)+1,$  \\[0.2cm]
        \quad $(m_1+1)(n_1+1)+2,\ldots$ &\quad $\ldots,N-m_{1}(n_{1}+1)-2,N-m_{1}(n_{1}+1),$  \\[0.2cm]
        & \quad $N-m_{1}(n_{1}+1)+1,\ldots,N-1,N+n_{1}$  \\[0.2cm]
   
  \hline \hline
\end{tabular}
\end{center}

\end{table}\par
%
%
\begin{table}[h!]

\caption{The states $\xi$ annihilated by the integrals $I_{-}$ in the $|N, \nu_{x} \rangle$ basis for the superintegrable systems $H_e$, $H_f$, and $H_g$ with $k_1=k_2=1$ and $m_1 \ge n_1$.}

\begin{center}
\begin{tabular}{ll}
  \hline\hline\\[-0.2cm]
    $N$ & $\nu_{x}$  \\[0.2cm]
  \hline\\[-0.2cm]
    $-m_{1}-n_{1}-1,-m_{1},-m_1+1,\ldots,-n_{1}-1$ & $-m_{1}-1$  \\[0.2cm]
    $-n_{1},-n_1+1,\ldots,0$ & $-m_{1}-1,N+n_{1}$  \\[0.2cm]
    $1,2,\ldots,m_{1}n_{1}-1$ & $-m_1-1,0,1,\ldots,N-1,N+n_{1}$  \\[0.2cm]
    $m_{1}n_{1}$ & $-m_{1}-1,0,1,\ldots,m_{1}n_{1}-1$ \\[0.2cm]
    $m_1n_1+1, m_1n_1+2, \ldots, (m_1+1)n_1$ & $-m_1-1, 0, 1, \ldots, N-1, N+n_1$ \\[0.2cm]
    $m_{1}n_{1}+n_{1}+1,m_1n_1+n_1+2,\ldots,m_{1}n_{1}+m_{1}$ & $-m_{1}-1,0,1,\ldots,(m_{1}+1)n_{1}-1,$  \\[0.2cm]
       & \quad $(m_{1}+1)n_{1}+1,\ldots,N-1,N+n_{1}$  \\[0.2cm]
    $m_{1}n_{1}+m_{1}+1,m_1n_1+m_1+2,\ldots,$ & $-m_{1}-1,0,1,\ldots,(m_{1}+1)n_{1}-1,$  \\[0.2cm]
       \quad $m_{1}n_{1}+m_{1}+n_{1}+1$ & \quad $(m_{1}+1)n_{1}+1,\ldots,N-1$  \\[0.2cm]
    $(m_{1}+1)(n_{1}+1)+1,(m_1+1)(n_1+1)+2,\ldots$ & $-m_{1}-1,0,1,\ldots,(m_{1}+1)n_{1}-1,$  \\[0.2cm]
       & \quad $(m_{1}+1)n_{1}+1,\ldots,(m_{1}+1)n_{1}+m_{1}$  \\[0.2cm]   
  \hline \hline
\end{tabular}
\end{center}

\end{table}
\par
\clearpage
%
%

\end{document}